\documentclass[twocolumn]{aastex62}

\received{January 26, 2018}
\submitjournal{ApJ}

\shorttitle{Hidden depths of planetary atmospheres}
\shortauthors{Yan B\'etr\'emieux \& Mark R. Swain}

\begin{document}

\title{The hidden depths of planetary atmospheres}

\correspondingauthor{Yan B\'etr\'emieux}
\email{Yan.Y.Betremieux@jpl.nasa.gov}

\author{Yan B\'etr\'emieux}
\affil{Jet Propulsion Laboratory, California Institute of Technology \\
4800 Oak Grove Drive, Pasadena, CA 91109, USA}

\author{Mark R. Swain}
\affil{Jet Propulsion Laboratory, California Institute of Technology \\
4800 Oak Grove Drive, Pasadena, CA 91109, USA}



\begin{abstract}

Atmospheric regions below a refractive boundary are hidden in limb observations. Refraction thus creates a gray continuum in the planet's transmission spectrum which can hide spectral features associated with sources of atmospheric opacity. We combine refractive theory with recent analytical advances describing the effects of surfaces and clouds on transmission spectra, to express the location of this boundary in atmospheric opacity space, both for atomic and molecular extinction, as well as collision-induced absorption. This allows one to quickly estimate how refraction affects spectral features in well-mixed atmospheres. We show that differences in the geometry of limb observations between solar system planets and exoplanets leads to different locations of this boundary, and that more than four extra scale heights of atmosphere are hidden in exoplanet transits compared to solar system observations of cold gas giants. We explore how the location of this refractive boundary in exoplanet transits changes in a well-mixed isothermal atmosphere with its temperature and composition, the spectral type of the planet's host star, and the size of the planet. We demonstrate that five extra scale heights of atmosphere are hidden in a terrestrial planet with a CO$_2$ atmosphere compared to a helium atmosphere, resulting in a flatter spectrum than from its smaller scale height alone. We provide results for a few exoplanets, notably those in the TRAPPIST-1 system, to help the scientific community estimate the impact of refraction on the size of spectral features without radiative transfer calculations, and thus help refine planned James Web Space Telescope observations.

\end{abstract}

\keywords{atmospheric effects -- methods: analytical -- methods: numerical -- planets and satellites: atmospheres -- radiative transfer.}



\section{Introduction}\label{intro}

Over the last decade, the reduction and interpretation of exoplanet transmission spectra have significantly improved. Much effort has gone into the development of data reduction techniques to remove the effects of instrument systematics affecting space observatories (see \citealt{Beichman_2014} for a review of the type and source of instrument systematics affecting Hubble Space Telescope, Kepler, and Spitzer observations), as well as correct for the effects of stellar variability from spots (e.g. see early work by \citealt{Pont_2008}; \citealt{Sing_2011}; \citealt{Pont_2013}; \citealt{McCullough_2014}), and many atmospheric parameter retrieval algorithms have been developed (e.g. \citealt{Irwin_2008}; \citealt{M_S_2009}; \citealt{B_S_2012}; \citealt{Line_2013}; \citealt{Waldmann_2015}; \citealt{M_M_2017}). Spectral signatures of molecular species are now routinely detected in transmission spectra (H$_2$O and CH$_4$ were first reported by \citealt{Tinetti_2007} and Swain, Vasisht \& Tinetti~2008, respectively, and TiO/VO only recently convincingly detected by \citealt{Evans_2016}) and so are the tangled signatures of H$_2$, clouds, and hazes (first reported by \citealt{Pont_2008} and identified by \citealt{Lecavelier_2008} in HD189733b's atmosphere), resulting in a few statistical studies of the abundance of water and the presence of clouds in hot exoplanets (\citealt{Sing_2016}; \citealt{Stevenson_2016}; \citealt{Iyer_2016}; \citealt{Barstow_2017}). However, the limited spectral coverage and signal-to-noise ratio of current observations, combined with the possible presence of hazes and clouds, pose a challenge to unambiguously retrieve the abundances of molecular species (see the sobering review by \citealt{Burrows_2014}). With the deployment of the James Web Space Telescope (JWST), good quality exoplanet spectra with a large spectral coverage may finally be obtained to overcome these limitations.

In anticipation, \citet{Beichman_2014} presented a comprehensive study and review of the various JWST instruments best suited for exoplanet science, as well as outlined the necessary steps which the scientific community must take to benefit from JWST's unique capabilities. Target lists\footnote{https://jwst-docs.stsci.edu/display/JSP \\ /JWST+GTO+Observation+Specifications} of JWST Garanteed Time Observations (GTO) have recently been released, and possible targets for the Early Release Science (ERS) Program have also been selected \citep{S_JWST_2016}. The expected performance of the various instruments covering the 0.6--30~\micron~spectral region have been modeled, and issues such as noise, instrument systematics, stellar spots, clouds, and hazes, impacting the accuracy of the retrieved atmospheric composition of various exoplanets, have been investigated (\citealt{Deming_2009}; \citealt{Barstow_2015}; \citealt{Greene_2016}; \citealt{Barstow_2016}; \citealt{B_I_2016}; \citealt{Arney_2017}; \citealt{Molliere_2017}) to help define the desired signal-to-noise and required integration time of JWST observations.

Among these studies, only the atmospheric models of \citet{Arney_2017} consider the effects of atmospheric refraction by the observed exoplanet on its transmission spectrum. However, this particular study focusses on hazy atmospheres where refraction is less likely to have an impact. All other studies ignore refractive effects, thus potentially underestimating the integration time required to detect various chemical species. Indeed, refraction can decrease the strength of absorption features by creating a grey continuum akin to an optically thick cloud deck (see \citealt{Munoz_2012}; \citealt{YB_LK_2013}, 2014; Misra, Meadows \& Crisp 2014 for terrestrial planets, and \citealt{YB_2016} for gas giants). The location of the critical boundary \citep{YB_LK_2014}, below which the atmosphere cannot be probed, depends both on the lensing power of the atmosphere and on the angular size of the host star viewed from the exoplanet.

One possible barrier to the widespread inclusion of refraction in exoplanet transmission models is that not only was it initially deemed unimportant (\citealt{Brown_2001}; \citealt{Hubbard_2001}), but its mathematical description is complex and requires numerical integrations. Indeed, \citet{YB_LK_2015} showed that the solution to the integral describing the dependence of the deflection of a ray with the atmospheric density at its grazing radius -- lowest region reached by a ray -- deviates at high densities (more than one amagat for Earth and temperate Jupiter-sized planets) from the simple analytical solution presented by \citet{B_C_1953} and still widely used today. This results in a thin refractive boundary layer across which the ray deflection goes from finite to infinite values as the grazing radius approaches a lower refractive boundary. Below this lower boundary, stellar radiation spirals into the planet until it is absorbed or scattered.

An observer's solution, which avoids computing exoplanet transmission spectra from first principle, is to build them up from the observed altitude-dependence of the limb transmission of atmospheres of solar system planets, correcting for differential refraction \citep{B_C_1953}. One method consists in observing the change in brightness of a moon as it enters its planet's shadow during a lunar eclipse to determine its transmission spectrum, as was done for Earth (\citealt{Palle_2009}; \citealt{VidalMadjar_2010}; \citealt{Munoz_2012}; \citealt{Arnold_2014}; \citealt{Yan_2015}). Another method is to observe the dimming of a star or our Sun as it is occulted by a planetary atmosphere, as was done with solar occultation data from the Cassini spacecraft, both for the atmospheres of Titan \citep{Robinson_2014} and Saturn \citep{Dalba_2015}. However, as we discuss in Section~\ref{probedregion}, the observational geometry of occultations and lunar eclipse observations is different from an exoplanet transit, and so is the relevant refractive boundary. Thus, inferring the transmission spectrum of an exoplanet from its solar system analog is not only challenging, but potentially misleading if the results are not interpreted properly in cases where refraction is important. 

The importance of refraction in a transmission spectrum can only be weighed against sources of atmospheric opacity, to determine to what degree the refractive continuum can reduce the strength or completely smother spectral features of interest. Although knowing the atmospheric pressure at the relevant refractive boundary (\citealt{YB_LK_2014}; \citealt{YB_2016}) is important to compare to the pressure location of the top of optically thick clouds or terrestrial surfaces, and determine which of these three types of `surface' \citep{YB_MS_2017} may be responsible for the observed continuum, it is not sufficiently informative to estimate the strength of absorption features without a detailed radiative transfer calculation. A more useful metric for back-of-the-envelope calculations is the concept of the `surface' cross-section introduced by \citet{YB_MS_2017} which translates the vertical location of a `surface' (actual surfaces, optically thick clouds, and refractive boundaries), given in term of its atmospheric density, into an effective opacity against which one can compare atmospheric opacities of interest.

The aim of this paper is to apply this novel concept to refractive effects in order to do several things.
First, to show that refraction substantially decreases spectral features of cold long-period giant exoplanets -- analogs to our Jovian planets -- and that, contrary to the claim by \citet{Dalba_2015}, it is probably not worthwhile to target these worlds for detailed atmospheric characterization. Second, to quantify how much refractive boundaries shift with the assumed bulk composition and temperature of exoplanet's atmospheres, with their size, and with the spectral class of their host star. Third, to provide to the scientific community a way to estimate, without radiative transfer calculations, the impact of refraction on the strength of absorption bands for a few exoplanets, notably the planets in the TRAPPIST-1 system, which will be targeted by JWST. As a by-product of these investigations, we also illustrate the effects of a non-isothermal temperature profile on transmission spectra and the importance of collision-induced absorption (CIA) in potentially defining the deepest atmospheric region that can be probed in hot exoplanets, as well as discuss the potential impact of collisional broadening on transmission spectra.

\section{Mathematical description}\label{math}

\subsection{Refractivity}\label{refractivity}

The bending of light due to refraction in a planetary atmosphere is essentially controlled by two quantities: the density scale height ($H$), and the refractivity ($\nu$) of the atmosphere, both evaluated at the grazing radius of a ray -- the deepest region reached by a ray. Although the refractivity of the atmosphere depends on its composition, it depends predominantly on its number density ($n$), which varies exponentially with altitude. We use the term `scale height' to signify the density scale height of the atmosphere, not the pressure scale height which we always label explicitly.

A starting point for computing the lensing power of an atmosphere is the tabulated refractivity (e.g. see Table~1 in \citealt{YB_LK_2015}) of different molecules ($\nu_{STP_j}$) measured at standard temperature and pressure (STP). Indeed, one first computes the STP refractivity of the atmosphere ($\nu_{STP}$) using
\begin{equation}\label{refraccompo}
\nu_{STP} = \sum_j f_j \nu_{STP_j} ,
\end{equation}
which depends on the sum of the STP refractivities of the individual species ($\nu_{STP_j}$) weighed by their abundances, quantified by their mole fraction ($f_j$). One then obtains the refractivity of the atmosphere at the desired density with
\begin{equation}\label{refrac}
\nu =  \left( \frac{n}{L_0} \right) \nu_{STP} ,
\end{equation}
where $L_0$ is the number density of the atmosphere at STP, also known as the Loschmidt constant. The unit of atmospheric density which equates $L_0$ is an amagat. It is clear from Equ.~\ref{refraccompo} that only the most abundant species contribute significantly to the atmospheric refractivity. Below a planet's homopause, the species making up the bulk of the atmosphere have a constant mole fraction with altitude so that the STP refractivity of the atmosphere is also constant with altitude. Although one might argue that this statement is not true of exoplanets with a high proportion of condensable species, such as `ocean-planets' \citep{Leger_2004}, the existence of these exoplanets is at this point speculative.

\subsection{Apparent versus actual probed region}\label{pvsact}

The path of a refracted ray through an atmosphere is described by an invariant which is equal to the ray's projected distance to the center of the planet with respect to the observer (Phinney \& Anderson 1968), also known as the ray's impact parameter ($b$). The ray's impact parameter is related to its grazing radius ($r$) by
\begin{equation}\label{raypath}
b = r (1 + \nu) .
\end{equation}
Without bending by refraction -- $\nu$ is equal to zero -- the impact parameter and the grazing radius are the same. Equation~\ref{raypath} can be rewritten to emphasize how much deeper a ray probes the atmosphere compared to what the impact parameter suggests when refraction is ignored.
\begin{equation}\label{diffimpgraz}
b - r =  r \nu 
\end{equation}
shows that the difference in altitude between the region perceived to be probed (given by $b$) and the actual region probed (given by $r$) increases with the refractivity -- and hence density -- of the atmosphere at the grazing radius, and thus with decreasing altitudes, as illustrated in Fig.~\ref{fig1}.

In an altitude region with a constant density scale height, the corresponding difference between the density probed ($n$) and the perceived probed density ($n^*$), or apparent density, is then given by 
\begin{equation}\label{perceived}
\frac{n^*}{n} = e^{-(b-r)/H} = e^{-(r \nu/H)} , 
\end{equation}
where the apparent density is the largest density reached if a ray was not bent by refraction. It is interesting to note that the factor $(r \nu/H)$ is a factor that also appears in the improved analytic expressions  \citep{YB_LK_2015} for the column abundance along a ray, as well as that for the ray deflection, and shown again in Section~\ref{probedregion}. Since this factor is a positive quantity, Equation~\ref{perceived} shows that the apparent density probed is always lower than the actual density probed. We want to stress that this factor is only a mathematical construct to map a change of altitude to a virtual change of probed density. In reality, rays do actually bear the signature of the actual density probed through the atmospheric extinction that they experience. Also, Equ.~\ref{perceived} does not hold for an atmosphere with a non-constant scale height. In that case, one must determine the change in density corresponding to the difference in altitude between the impact parameter and the grazing radius from the density altitude profile of the atmosphere.

\subsection{Lower boundary}\label{lowb}

The lowest region that stellar radiation can pass through an atmosphere without spiraling into the planet is the 
lower refractive boundary. This lower boundary (represented with the $lb$ subscript throughout the paper) occurs at an altitude which satisfies the condition
\begin{equation}\label{lowerbound}
\left( \frac{\nu_{lb}}{1+\nu_{lb}} \right) \frac{r_{lb}}{H_{lb}} = 1  ,
\end{equation} 
as derived by \citet{YB_LK_2015}. If the atmosphere is thin compared to the planetary radius ($R_P$), then $r_{lb} \approx R_P$, and one can then solve for the atmospheric refractivity at this lower boundary with 
\begin{equation}\label{refraclb}
\nu_{lb} = \frac{1}{(r_{lb}/H_{lb})-1} \approx H_{lb}/R_P ,
\end{equation}
which is then much smaller than one. In this approximation, we can see that the condition describing the location of the lower boundary (Equ.~\ref{lowerbound}) is simply $(r\nu/H)_{lb}~\approx~1$. Knowing $\nu_{lb}$, we can then use Equ.~\ref{refrac}, with the assumed bulk composition of the atmosphere, to determine the density where the lower boundary is located.

\begin{figure}
\includegraphics[scale=0.50]{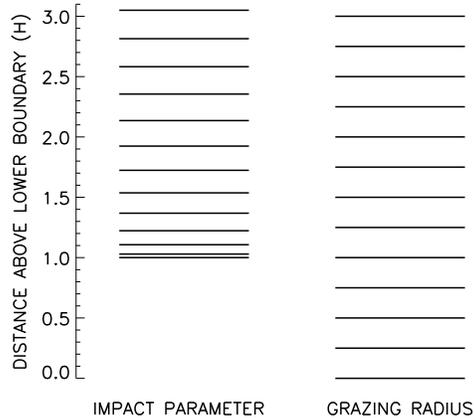}
\caption{Illustration of the difference between the lowest altitude probed (or grazing radius) by a ray, and the apparent altitude probed (or impact parameter) seen by an observer outside the atmosphere during limb observations. On the right, the rays are evenly spaced in altitude (expressed in units of scale height) above the location of the lower refractive boundary (see Section~\ref{lowb}). The apparent altitudes probed, on the left, are always located at a higher altitude than the actual altitudes probed, and the effective scale height of the atmosphere (see Section~\ref{effscl}) shrinks as rays graze closer to the lower refractive boundary.
\label{fig1}}
\end{figure}

Since the difference in altitude between the apparent and actual probed region increases with density (Equ.~\ref{diffimpgraz}), the maximum difference occurs at this lower boundary. With the help of Equ.~\ref{lowerbound}, we can see that the difference between the apparent and actual altitude of the lower boundary is given by
\begin{equation}\label{maxdiff}
(b_{lb} - r_{lb}) =  r_{lb} \nu_{lb}  = H_{lb} (1+\nu_{lb}) .
\end{equation}
Since $\nu_{lb}$ is typically much smaller than one, this difference is about one density scale height, as illustrated in Fig.~\ref{fig1}. This expression is exact even in an atmosphere with a non-constant scale height. 

\subsection{Effective density scale height}\label{effscl}

\begin{figure}
\includegraphics[scale=0.5]{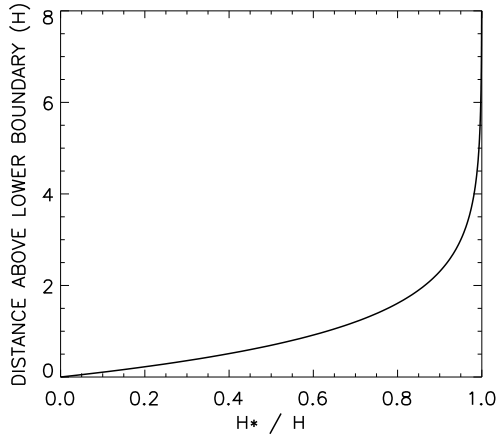}
\caption{Ratio of the effective scale height ($H^*$) perceived by an observer outside the atmosphere to the actual scale height ($H$) of the probed atmospheric region as a function of its altitude above the lower refractive boundary (expressed in units of scale height). 
\label{fig2}}
\end{figure}

The fact that the difference between the apparent and actual altitude probed increases with density implies that the density scale height appears to change to an external observer even in an atmosphere where the scale height is constant with altitude. Fig.~\ref{fig1} illustrates how rays that are evenly spread in grazing radius -- corresponding to intervals of constant ratios of probed densities -- map into their respective impact parameters for an atmosphere with a constant scale height. Fig.~\ref{fig1} clearly shows that constant grazing radius intervals are mapped to smaller impact parameter interval the closer a ray gets to the lower refractive boundary. Since the probed densities have not actually changed, the same ratio of density occurs over a smaller altitude interval, and the scale height appears to have decreased. The effective, or apparent, density scale height ($H^*$) is given by 
\begin{equation}
 \frac{H^*}{H} = \frac{db}{dr} = (1 + \nu) - \left(\frac{r\nu}{H}\right)
\end{equation}
\citep{YB_LK_2015}, and shown in Fig.~\ref{fig2}. Six scale height above the lower boundary, the effective scale height is essentially the same as the actual scale height. However, two scale heights from the lower boundary, the ratio of the effective to the actual scale height drops to less than 0.9, and decreases dramatically to zero in the bottom scale height. It can be shown that the dominant term describing the curve in Fig.~\ref{fig2} is simply
\begin{equation}
 \frac{H^*}{H} \approx 1 - e^{-(r - r_{lb})/H} 
\end{equation}
for an atmosphere with a constant scale height. The fact that the effective density scale height decreases as the lower boundary is approached implies that the associated spectral features which probe these atmospheric regions also decrease in size \citep{YB_2016}.

\subsection{Atmospheric regions probed by refraction}\label{probedregion}

A thin refractive boundary layer which distorts the apparent scale height of the atmosphere occurs only at relatively high densities, as shown for Jovian atmospheres of different temperatures in Table~3 of \citealt{YB_2016}. Whereas this boundary could be observed in planets with thick atmospheres in our Solar System in spectral regions of low opacity, the observational geometry of transiting exoplanets can hide these deeper regions \citep{Sidis_Sari_2010} because it creates a refractive boundary located at a higher altitude.

This critical boundary \citep{YB_LK_2014} occurs at an altitude where the atmospheric density deflects light from the stellar limb opposite the planetary limb toward the observer (see their Fig.~2). This critical deflection ($\omega_{c}$), in turn, depends on the planet-star geometry and is given by 
\begin{equation} \label{critdef}
\sin\omega_{c} = \frac{r_c + R_{\star}}{d} ,
\end{equation}
where $r_c$ is the critical grazing radius ($\approx R_P$), $R_{\star}$ is the stellar radius, and $d$ is the planet-star distance. 

Atmospheric regions below the critical boundary cannot be probed because they are not back-illuminated by the host star from the observer's viewpoint, and thus appear dark. Indeed, the atmosphere acts as a lens which severely distorts the imaged stellar disc. Although the imaged stellar disc varies substantially during the course of a planetary transit (\citealt{Sidis_Sari_2010}; \citealt{Munoz_Mills_2012}; \citealt{MMC_2014}), one can use the simple geometry when the planet occults the center of its star to determine the average effects of refraction during the transit. Indeed, the actual locations of the imaged stellar limb on opposite sides of the planetary limb straddle the one provided by this simple picture. Furthermore, all radiative transfer codes, used in retrieval algorithms to interpret transit spectra, assume this simple geometry.

Determining the altitude where the critical boundary occurs must be done numerically because there exist no analytic expression for the ray deflection at high densities. Indeed, the formula for the ray deflection ($\omega$) in a constant scale height atmosphere can be expressed as a Taylor series by
\begin{equation}\label{deflection}
 \omega = \sqrt{\frac{2\pi r}{H}} \nu \left[1 + \sum_{j = 0}^{\infty} D_j \left( \frac{r\nu}{H} \right)^j \right] .
\end{equation}
A similar type of behavior exists for the column-integrated abundance ($N$) along the ray, namely
\begin{equation}\label{column}
 N = \sqrt{2\pi r H} n \left[1 + \sum_{j = 0}^{\infty} C_j \left( \frac{r\nu}{H} \right)^j \right] .
\end{equation}
Here, $C_j$ and $D_j$ are coefficients for the column abundance and deflection, respectively, of which only the first few are determined \citep{YB_LK_2015}. Both of these expressions are expected to diverge at the lower boundary. The factors outside the parenthesis of both expressions are well-known formulae (e.g. see \citealt{B_C_1953}, \citealt{S_H_1990}, and \citealt{Fortney_2005}) valid at low densities. Throughout the paper, we use the new numerical ray tracing method described by \citet{YB_LK_2015} to determine $\omega$ and $N$ through the atmosphere, rather than use Equ.~\ref{deflection} and \ref{column}. Percentage differences of the values of $\omega$ and $N$ from the numerical method, as well as from Equ.~\ref{deflection} and \ref{column}, compared to the well-known simple formulae, have already been shown in Fig.~6 and 7 of \citet{YB_LK_2015} for a temperate Jovian and an Earth-like planet.

\citet{YB_LK_2014} showed that for low densities, the critical density ($n_c$) -- atmospheric number density at the critical boundary -- is given by
\begin{equation}\label{simpcritn}
\left( \frac{n_c}{L_0} \right) = \frac{\omega_c}{\nu_{STP}} \sqrt{\frac{H}{2\pi r}} ,
\end{equation}
where the ratio inside the parenthesis on the left-hand side of the equation is a number density expressed in units of amagat. This simple recipe, which ignores the concept of the lower boundary, cannot be applied to hot Jupiters because the critical boundary is located at high densities where Equ.~\ref{simpcritn} breaks down. Indeed, numerical results show that the critical boundary of a 1200~K isothermal hot Jupiter, orbiting a star cooler than an F0 spectral class, is located within one scale height of the lower boundary (see Table~3 of \citealt{YB_2016}), well inside the refractive boundary layer. Using Equ.~\ref{simpcritn} when 
\begin{equation}\label{conditwrong}
\omega_c \gtrsim  \sqrt{\frac{2\pi H}{r}} ,
\end{equation}
yields densities larger than at the lower boundary, which is unphysical.

Since the ray deflection diverges at the lower boundary, the critical boundary, which causes a finite ray deflection, is always located above it and defines the atmospheric region which can be probed in the course of an exoplanet's transit. Limb observations of a planetary atmosphere from an orbiting spacecraft, via solar or stellar occultations, are not similarly restricted because there will always be an atmospheric layer which bends the light sufficiently to reach the spacecraft. The only issue is whether opacity and differential refraction have sufficiently decreased the transmitted flux to the point of non-detection (see review of occultations by \citealt{S_H_1990}). Observing the brightness of lunar surfaces during a lunar eclipse is potentially more powerful as it provides a simultaneous coverage of the transmission of many atmospheric layers. The drawback is that scattered light contribution now comes from the entire limb of the planet. Both methods can in principle probe an atmosphere to its lower boundary.

The difference in altitude (or density) between the lowest atmospheric regions probed by exoplanet transmission spectroscopy (critical boundary), and solar/stellar occultations and lunar eclipses (lower boundary), can be problematic when trying to infer the transmission spectrum of an exoplanet from transmission data of a solar system analog, as was done for Titan \citep{Robinson_2014} and Saturn \citep{Dalba_2015}. Indeed, the greater the difference in altitude, the more the spectral features in the exoplanet transit will be decreased compared to the spectrum built-up from transmission data of its solar system analog. If the lower boundary and the critical boundary are widely separated, it is entirely possible for refraction to hardly impact the spectrum of solar system planets, and to cut down most of the absorption features in an identical exoplanet observed in transit to the point of resulting in a nearly flat spectrum.

\subsection{Refraction and atmospheric opacities}\label{opacityref}

To determine if refraction plays an important role in shaping the transmission spectrum of planetary atmospheres, one must determine the location of the refractive continuum with respect to spectral features of interest. Until recently, the few analytical formalisms (\citealt{Lecavelier_2008}; \citealt{dW_S_2013}) attempting to explain what shapes exoplanet transmission spectra could only do so for clear atmospheres. However, \citet{YB_MS_2017} developed an analytical formalism which explains how `surfaces' (surfaces, thick cloud decks, and refractive boundaries) decrease the contrast of absorption features in transmission spectra. At the heart of their formalism is the only published solution (see Equ.~27 in \citealt{YB_MS_2017}) to the integral describing the effective radius of a transiting exoplanet whose atmosphere is completely opaque below a `surface'. Their formalism is the only one that gives the expected result that as the optical thickness of the atmosphere decreases, the effective radius of the exoplanet approaches asymptotically that of the location of the `surface' without going below it. Indeed, the effective radius of a terrestrial planet without an atmosphere is simply the radius of the planet -- the location of its surface.

\citet{YB_MS_2017} also showed that, in lieu of a detailed radiative transfer calculation, one can estimate the effective atmospheric thickness ($h$) -- atmospheric contribution to the transmission spectrum -- by comparing the mean atmospheric cross-section ($\sigma$) to a `surface' cross-section ($\sigma_s$) with
\begin{equation}\label{spectralmod}
h / H = \ln(\sigma / \sigma_s) , 
\end{equation}
and setting negative results to zero. This simple method produces at most half of a scale height error (see Fig.~5 in \citealt{YB_MS_2017}) when $\sigma = \sigma_s$, i.e. when the atmospheric contribution is small. This is equivalent to the ad hoc method used to incorporate the effects of clouds in a few past analyses (\citealt{Berta_2012}; \citealt{Sing_2015}, 2016), except that it is here done in opacity space, and that the errors associated with that method are now known. Equation~\ref{spectralmod} assumes that the atmosphere has not only a constant scale height, but is also well-mixed so that the mole fraction of species ($f_j$) are constant with altitude. The mean atmospheric cross-section, which is given by 
\begin{equation}\label{atmcross}
\sigma = \sum_j f_j \sigma_j , 
\end{equation}
is thus also constant with altitude, and is sufficient to characterize the opacity of an atmosphere. When the extinction in a spectral region is due predominantly to one species, the mean atmospheric cross-section is simply given by the product of the abundance and the extinction cross-section of that species ($\sigma \approx f_j \sigma_j$). The `surface' cross-section is computed with 
\begin{equation}\label{surfcross}
\sigma_s = \frac{e^{-\gamma_{EM}}}{\sqrt{2\pi b_s H} n_s} , 
\end{equation}
where $\gamma_{EM}$ is the Euler-Mascheroni constant ($\approx 0.577$), while $b_s$ and $n_s$ are the impact parameter (Equ.~\ref{raypath}) and the atmosphere number density, respectively, at the `surface'. 

Although the well-mixed and isothermal assumptions are not valid over all altitudes in a planetary atmospheres, they are usually deemed sufficient when it comes to the retrieval of atmospheric parameters from transmission spectra. 
Indeed, most exoplanet atmosphere parameter retrieval algorithms (e.g. \citealt{M_S_2009}; \citealt{B_S_2012}; \citealt{Waldmann_2015}; \citealt{M_M_2017}; \citealt{Barstow_2017}) assume that atmospheres are well-mixed in the range of altitudes probed by transmission spectra. Regarding the effects of temperature on transmission spectra, it is clear that a vertical change in temperature changes the scale height, which in turn should impact the size of spectral features. In Section~\ref{tempeffects}, we discuss further the effects of a non-isothermal temperature profile on transmission spectra, but for the present discussion we assume that the isothermal profile assumption holds over the altitude range probed by transmission spectroscopy. 

We also note that the assumption of a constant mean cross-section with altitude is not valid in the denser atmospheric regions where collisional broadening becomes important (see the Appendix for a detailed discussion about the impact of collisional broadening on transmission spectra). However, not only do prominent absorption bands in exoplanet transmission spectra typically probe lower pressures where cross-sections are constant with altitude, but it is impossible to ascertain the impact of collisional broadening in a hot Jupiter's atmosphere because broadening parameters by H$_2$ and He are non-existent except for a few molecules (\citealt{Wilzewski_2016}; \citealt{Barton_2017}). Even for air as a broadening agent, parameters are usually available only around 300~K \citep{H_M_2016}. Given this lack of data, one can use our simple method because cross-sections can only be computed for zero pressure, and are thus independent of altitude in an isothermal atmosphere.

Equation~\ref{spectralmod}, \ref{atmcross}, and \ref{surfcross} are valid for Rayleigh scattering, or atomic and molecular absorption. Since the absorption coefficient of collision-induced absorption (CIA) is proportional to the square of the density, the relevant equations for CIA are slightly different \citep{YB_MS_2017}, and are given instead by 
\begin{equation}\label{spectralmodCIA}
(h / H)_{CIA} = \frac{1}{2} \ln(k / k_s) , 
\end{equation}
\begin{equation}\label{atmcrossCIA}
k = \sum_{i,j} f_i f_j k_{i,j} , 
\end{equation}
and 
\begin{equation}\label{surfcrossCIA}
k_s = \frac{e^{-\gamma_{EM}}}{\sqrt{\pi b_s H} n^2_s} ,
\end{equation}
where $k$ and $k_s$ are the mean atmospheric cross-section and the `surface' cross section for CIA, respectively, while $k_{i,j}$ is the CIA cross-section between species $i$ and $j$. Simply put, the relevant length scale that determines the size of CIA features is in fact half the scale height (\citealt{dW_S_2013}; \citealt{YB_MS_2017}). Throughout the paper, we use the subscript $c$ and $lb$ instead of $s$ with both types of cross-section ($\sigma$ and $k$) to identify them as `surface' cross-section at the critical boundary and lower boundary, respectively.

What number density should one use in Equ.~\ref{surfcross}, and \ref{surfcrossCIA} to determine, to first order, the various `surface' cross-sections of the refractive boundary? One should use the apparent density (See Section~\ref{pvsact}) of the refractive boundary appropriate for the type of observations: apparent critical boundary ($n_s = n_c^*$) for exoplanet transits, and apparent lower boundary ($n_s = n_{lb}^*$) for solar/stellar occultations and lunar eclipses observations (see Section~\ref{probedregion}). Indeed, the apparent density essentially translates the refractive case into a non-refractive framework when it comes to the location of the ray. The concept of the `surface' cross-section then determines, inside a non-refractive framework, the required opacity of an homogeneous clear isothermal atmosphere which causes the effective radius of the exoplanet to be located at the `surface'.  

\section{Solar system planets}\label{solplanets}

\subsection{Vertical density profile}\label{densprof}

To illustrate how refraction impacts in a different way limb observations of solar system planets and exoplanet transits, we determine the location of refractive boundaries for solar system planets with thick atmospheres (Venus, Jupiter, Saturn, Uranus, and Neptune) for both observational geometries. We first use the ideal gas law on published temperature-pressure altitude profiles (see Table~\ref{T-Pprof}) for the various planets considered to determine the number density of their atmosphere as a function of altitude. We then interpolate densities on a finer altitude grid assuming that the density follows an exponential law with a constant density scale height between altitudes with known values. Temperatures are interpolated linearly with altitude. The altitude spacing on the fine grid is set at 1/3000 of the density scale height at the bottom of the published atmosphere. The temperature-pressure profiles listed in Table~\ref{T-Pprof} come from the Galileo probe (Jupiter), a combination of American and Russian probes and orbiters (Venus), and occultation of radio signals from the Voyager~2 spacecraft by the planet's atmosphere (Saturn, Uranus, Neptune).

The detailed temperature-pressure altitude profile from the Voyager~2 radio occultations (\citealt{Lindal_1985}; \citealt{Lindal_1987}; \citealt{Lindal_1992}) was retrieved simultaneously with the composition of the atmosphere. If one uses a different atmospheric composition than these published compositions, one must modify the temperature and pressure according to a scaling law (e.g. \citealt{Lindal_1985}; \citealt{C_G_2000}), given by
\begin{equation}\label{scaleT}
\frac{T}{T_0} = \frac{\overline{m}}{\overline{m}_0}
\end{equation}
and
\begin{equation}\label{scaleP}
\frac{P}{P_0} = \left( \frac{\overline{m}}{\overline{m}_0} \right) \left( \frac{\nu_0}{\nu} \right) ,
\end{equation}
where the subscript $0$ marks the old values of the temperature ($T$), pressure ($P$), mean molecular weight of the atmosphere ($\overline{m}$), and the refractivity ($\nu$). We use more recent determinations of the composition of the atmospheres of Saturn and Uranus (see the compositions in Table~\ref{solsysplanets}), and Table~\ref{T-Pprof} lists the scaling factors for the temperature and the pressure that we apply to the published profiles prior to computing the densities.

\begin{table}
 \centering
  \caption{Published temperature-pressure altitude profiles used in this work, and the corrective factor applied to the temperature and pressure to account for an atmosphere composition different from the referenced source (see Section~\ref{densprof}).}
  \begin{tabular}{@{}llcc@{}}
  \hline
   Planet    &    Reference  &  ($T/T_0$) &  ($P/P_0$) \\
 \hline
 Venus & \citet{Seiff_1985} & - & - \\
 Jupiter & \citet{Seiff_1998} & - & - \\
 Saturn & \citet{Lindal_1985} & 1.084 & 1.123 \\
 Uranus & \citet{Lindal_1987} & 1.141 & 1.075 \\
 Neptune & \citet{L_F_1998}, & 1 & 1 \\
   &  and \citet{Lindal_1992} &   &   \\
\hline
\end{tabular}\label{T-Pprof}
\end{table}

Although we use the same composition and temperature-pressure profile for Neptune as originally published by \citet{Lindal_1992}, the model atmosphere of Neptune present its own set of challenges, namely that the altitudes at which a given pressure and temperature occur are not specified. Only in \citet{L_F_1998}, based on \citet{Lindal_1992}, are the altitudes specified but only for pressures of 1~bar or lower. We compute the various altitude interval between adjacent points for pressures greater than 1~bar by assuming that the temperature and pressure follow an adiabatic law with a constant ratio of specific heat (see Section~\ref{tempeffects}), which is itself determined empirically from the temperature-pressure power law index in each altitude interval of the model atmosphere.

\begin{table*}
 \centering
 \begin{minipage}{130mm}
  \caption{Refractive boundaries of solar system planets with thick atmospheres}
  \begin{tabular}{@{}lccccc@{}}
  \hline
     Parameters     & Venus & Jupiter &  Saturn & Uranus & Neptune \\
 \hline
     Input~\footnote{Source: planetary fact sheets available on the NASA Space Science Data Coordinated Archive (NSSDCA) website (https://nssdc.gsfc.nasa.gov/).} &    &    &    &    &   \\
 \hline
  Volumetric radius (km) & 6051.8 & 69911 & 58232 & 25362  & 24622 \\
  Mass ($10^{24}$ kg)   & 4.8676 & 1898.3  & 568.36 & 86.816  & 102.42 \\
  Semi-major axis (AU)  & 0.7233 &  5.204 & 9.752 & 19.201 & 30.047 \\
  \hline
 Atmospheric Composition~\footnote{Source: \citet{L_F_1998}, unless specified otherwise. Quantities are mole fractions.}   &  &   &  &   & \\
 H$_2$   & - & 0.8620  & 0.8766~\footnote{Adjusted so that the sum of all mole fractions is 1.} &  0.825 & 0.80\\
 He   & - &  0.1356 & 0.1189~\footnote{Source: \citet{C_G_2000}.} & 0.152  & 0.19 \\
 CH$_4$   & - &  1.81$\times10^{-3}$ & 4.5$\times10^{-3}$ & 0.023 & 0.01 \\
 CO$_2$   & 0.965 &  - & - &  - & - \\
 N$_2$   & 0.035 & -  & - &  - & - \\
  \hline
 Derived~\footnote{Lower boundaries for Saturn, Uranus, and Neptune are located deeper than the highest pressures in their published temperature-pressure profile (Table~\ref{T-Pprof}). Quantities related to the location of the lower boundary usually show two values which bracket the range of possible solutions: one, marked with a  $^{\star}$, which assumes a constant $(r/H)$ ratio below the deepest layer, and another one which assumes an adiabatic profile. For Neptune, both solutions give about the same value.} &              &      &      &      &        \\
\hline
$P_c^*$ (mbar) &  154  &   24.7   &  15.3   &   6.5   &  4.9      \\
$P_c$ (mbar)  &   168  &   25.2   &  15.4   &   6.5   &  4.9      \\
(r$\nu$/H)$_c$ & 0.081 & 0.025 & 0.0091 & 0.0038 & 0.0024 \\
$P_{lb}^*$ (bar) &     2.3    & 1.9     &  $2.0^{\star} - 4.6$  &   $3.0^{\star} - 36$    &  $3.2^{\star} - 2.4$   \\
$P_{lb}$ (bar) &     8.7   &  12.2    &   $5.6^{\star} - 25$  & $8.1^{\star} - 460$    &  $8.8^{\star} - 13$     \\
(r$\nu$/H)$_{s}~$\footnote{Evaluated at the bottom of the atmosphere model listed in Table~\ref{T-Pprof}. The lower refractive boundary is located within the extent of the atmosphere model when this value is greater than one.} & 4.8 & 1.3 & 0.26 & 0.30 & 0.71 \\
$\sigma_c$ ($10^{-27}$ cm$^2$ molec$^{-1}$)  & 2.7   &  1.5   &  1.6   & 5.0    &   8.3   \\
$\sigma_{lb}$ ($10^{-29}$ cm$^2$ molec$^{-1}$)  & 30   &  2.5   &  $0.62 - 1.1^{\star}$   & $0.14 - 1.1^{\star}$    &  1.2   \\
$k_c$ ($10^{-45}$ cm$^5$ molec$^{-2}$)           &  0.88  &  1.6   &  2.1   &  11   &    29   \\
$k_{lb}$ ($10^{-49}$ cm$^5$ molec$^{-2}$)           &  102  &  5.3   &  $0.56 - 1.7^{\star}$   &  $0.025 - 0.81^{\star}$   &    1.0 \\
$\ln(\sigma_c/\sigma_{lb})~$\footnote{Difference in effective atmospheric thickness between the two refractive continuums, expressed in density scale height.} & 2.2 & 4.1 & $5.0^{\star} - 5.5$ & $6.1^{\star} - 8.2$ &  6.5  \\
\hline
\end{tabular}\label{solsysplanets}
\end{minipage}
\end{table*}

\subsection{Location of refractive boundaries and their impact on spectral features}\label{locrefrac}

For each planet, we compute the deflection for 80 rays evenly spread in grazing altitude from the refractive boundary upward across the vertical extent of its atmosphere with MAKEXOSHELL, which uses the new ray tracing algorithms described by \citet{YB_LK_2015} and is accurate over a large dynamical range of densities. This Fortran code numerically integrates along the path of rays on the fine altitude grid, starting from the density profile of the atmosphere and its composition. The required refractivity is determined from the density and composition of the atmosphere with Equ.~\ref{refrac}, while the density scale height is derived directly from the altitude dependence of the density profile. Early versions of MAKEXOSHELL (\citealt{YB_LK_2015}; \citealt{YB_2016}) required human intervention at a few intermediate steps. Given the aim of this paper, which requires computing the location of the refractive boundary over some part of parameter space, we have substantially improved MAKEXOSHELL such that all of the steps described in Sections~\ref{densprof} and~\ref{locrefrac} (except for Equ.~\ref{scaleT} and~\ref{scaleP} which modify the input temperature-pressure profile) are now fully automated.

Most of the improvement pertains to the automatic determination of the location of the refractive boundary itself and the subsequent set-up of the rays, which is done in the following fashion. We first determine the location of the lower boundary by evaluating the left-hand side of Equ.~\ref{lowerbound} on the fine altitude grid and finding the lowest altitude where that value is just barely less than one. We set-up the rays using this boundary as the location of the deepest ray, and compute the deflection of each ray across the limb. We then interpolate the deflection values onto the fine altitude grid by assuming that the ray deflection follows an exponential with a constant scale height between the computed values. We also compute on the fine altitude grid the critical deflection of the planet-star system (Equ.~\ref{critdef}) using the inputs from Table~\ref{solsysplanets} and the radius of the Sun ($6.96 \times 10^{5}$~km). We choose for our critical boundary the highest altitude where the ray deflection is still larger than the critical deflection. We then determine, from their grazing altitudes, the impact parameter of both boundaries (Equ.~\ref{raypath}). From a planet's temperature-pressure altitude profile, we can translate grazing altitudes and impact parameters to actual and apparent densities, as well as pressures, by interpolation rather than from Equ.~\ref{perceived}, which is valid only for an isothermal atmosphere. From the densities, we finally compute the various `surface' cross-sections (Equ.~\ref{surfcross} and~\ref{surfcrossCIA}). 

The results from our calculations are displayed in Table~\ref{solsysplanets}. As previously demonstrated by \citet{YB_LK_2014} and \citet{YB_2016}, the critical pressure is smaller for a planet further away from its star. We can also see that at the low pressures of the critical boundary, the $(r\nu/H)_c$ factor is much smaller than one, and the actual and apparent critical pressures are similar. For our Jovian planets, the critical pressure is located at low pressures, and refraction hides their tropospheric clouds. For Venus, the situation is reversed, and the critical boundary is located below the optically thick clouds of Venus whose top is situated at an average pressure of 37~mbar (or an average altitude of 70~km), as previously pointed out by \citet{Munoz_Mills_2012}. Thus, for a Venus analog, dominated by hazes and clouds, refraction plays very little role on its transmission spectrum.

For solar system planets, lower boundaries are located much deeper in the atmosphere than critical boundaries. For Venus and Jupiter, where in-situ measurements are available, we determine that the lower boundary occurs at pressures on the order of 10~bar, deep in their troposphere, which translates to apparent pressures of about 2~bar. For Saturn, Uranus, and Neptune, the model atmospheres retrieved from Voyager~2 radio occultations do not go deep enough to probe the lower boundary. Indeed, the $(r\nu/H)_s$ factor at the bottom of the model atmosphere is for these three planets less than one, and we cannot use the method described earlier to determine the location of the lower boundary. Instead, we consider two possible solutions which bracket the possible range of temperature gradients in these deeper regions. One solution assumes that the temperature-pressure profile follows an adiabatic law. In this case, we extrapolate the temperature profile to larger pressures using the same ratio of specific heat as in the deepest region of the model atmosphere, and then proceed as before. The other solution, almost equivalent to an isothermal profile, assumes that $(r/H)$ is constant downward from the bottom of the model atmosphere. In this approximation, the density at the lower boundary is simply given by $\nu_{lb} \approx H_s/r_s$ (see also Equ.~\ref{refraclb}), where quantities are evaluated at the bottom of the measured atmosphere. It is interesting to note that in the case of Neptune, even though the two solutions of the pressure of the lower boundary differs, the `surface' cross-section is identical to two significant digits. This clearly shows that knowing the deepest pressure probed by refraction is not enough to understand what happens to absorption features because the density scale height of the atmosphere also matters.

The large difference between the critical and the lower boundary, shown in Table~\ref{solsysplanets}, can potentially be a problem when trying to infer the transmission spectrum of an exoplanet analog to a solar system planet, as we already pointed out in Section~\ref{probedregion}. For example, \citet{Dalba_2015} recently used solar occultations viewed from the Cassini spacecraft to build-up a transmission spectrum of Saturn's atmosphere. They claimed that refraction was responsible for the observed continuum, which they estimated was located at a pressure of $1.0\pm0.5$~bar. They further showed that the continuum had been shifted upward by refraction from its actual location of about 2~bar, resulting in about a 10~part-per-million (ppm) increase in the transit depth of the continuum, and a corresponding decrease of the size of spectral features to a maximum size of about 90~ppm. They concluded that since cold planets such as Saturn could show as much as 90~ppm of spectral modulation, that it demonstrated that transmission spectroscopy was a viable technique to study cold long-period exoplanets analog to Saturn.

Our results demonstrate that this is not the case. For a solar occultation, the relevant refractive boundary is the lower boundary, which their observations support. Indeed, their continuum is located at a pressure of about 1~bar, a slightly lower pressure than the possible range of pressures of 2--4.6~bar of the apparent lower boundary of Saturn.  We can only deduce that the continuum in their spectrum must be due not only to refraction but also to some opacity from clouds and gas extinction. However, we determine that the critical boundary of Saturn is located at about 15~mbar, a much lower pressure than the claimed location of their continuum. We can estimate the change in transit depth due to the change in the location of the continuum. We compute that Saturn's critical boundary is located about 150~km ($\Delta z$) above its 1~bar pressure level, which results in a change of transit depth ($2R_P\Delta z/R_\star^2$) of about 36 ppm. Fig.~7 in \citet{Dalba_2015} shows five prominent spectral features from 1 to 5~\micron, three of which have peaks about 45~ppm above the continuum, and the other two stand at 55~pm and 90~ppm above the continuum. We can see that a 36~ppm upward shift of the continuum severely decreases the size of all spectral features except for the largest one which is still reduced by about 40\% in size. Hence, the transmission spectrum of a cold long-period exoplanet is probably nearly featureless except for the strongest features, contrary to the conclusions reached by \citet{Dalba_2015}.

A quick way to determine how much spectral features are reduced by refraction, which does not require a detailed radiative transfer computation, is through the `surface' cross-sections which we provide for both refractive boundaries in Table~\ref{solsysplanets}. We can compare the mean cross-section of the atmosphere to the `surface' cross-section which essentially defines the minimum mean cross-section of the atmosphere which produces spectral features. Any spectral features below that `surface' cross-section will be severely decreased as demonstrated in Fig.~6 in \citet{YB_MS_2017}. This comparison is trivial in the case when one species is predominantly responsible for the opacity in a given spectral region, as only the abundance and the cross-section of that species matter (Equ.~\ref{atmcross}). For CIA, we can hold the exact same type of reasoning, except that the mean cross-section depends on the product of the abundances of two responsible species (Equ.~\ref{atmcrossCIA}), and that the comparison must be done to the `surface' CIA cross-section. 

As an example, let's suppose that the extinction cross-section of a species of interest ranges from $10^{-25}$ to $10^{-21}$~cm$^2$~molec$^{-1}$ inside an observed spectral range. The maximum spectral modulation -- change of effective thickness, expressed in scale height or transit depth -- from a single species is simply given by the natural logarithm of the ratio of the maximum and minimum values of the cross-section (see the detailed discussion in Section~3.3 and~3.4 in \citealt{YB_MS_2017}), which is here 9.2 scale height. For CIA, the same ratio of cross-sections would lead to half the spectral modulation (Equ.~\ref{spectralmodCIA}). The `surface' cross-section of Uranus is $5 \times 10^{-27}$~cm$^2$~molec$^{-1}$. If our species of interest makes up the bulk of the atmosphere and $f \approx 1$, then no part of its cross-section is hidden by refraction. If the abundance of our species is $10^{-4}$, then the mean cross-section ranges from $10^{-29}$ to $10^{-25}$~cm$^2$~molec$^{-1}$, and the maximum spectral modulation which can be observed is given by $\ln(10^{-25}/5 \times 10^{-27})$, or about 3 scale heights. In this particular example, the maximum size of the spectral feature of our species has been reduced by more than a factor of three from 9.2 to 3 scale heights. If the abundance of our species is only $10^{-6}$, no part of the mean cross-section has values above the `surface' cross-section, and that species is to first-order well hidden by refraction. Thus, the abundance of a species controls the size of its spectral features above the refractive continuum \citep{YB_2016}, and above the continuum created by clouds and surfaces \citep{YB_MS_2017}. 

We can evaluate how many more scale heights of atmosphere does the refractive continuum in an exoplanet transmission spectrum hides compared to a spectrum constructed from solar system observations. This is done by computing $\ln(\sigma_c/\sigma_{lb})$, whose values for the different planets are listed at the bottom of Table~\ref{solsysplanets}. This difference is at least four scale heights for the giant planets and generally increases as the orbital distance increases. Uranus displays a much larger value for its solution of the adiabatic extrapolation because the lower boundary is located at a high pressure of 460 bar, a direct consequence of the steep temperature gradient at the bottom of its model atmosphere. It is clear from Table~\ref{solsysplanets} that the shift in the refractive continuum is large for exoplanets which are analogs to the cold long-period gas giants in our solar system, and that transmission spectra inferred from solar system observations overestimates the size of spectral features compared to exoplanet transits.

\subsection{Effective radius with opacity - Jupiter as an example}\label{Jupiter}

The first-order method described at the end of Section~\ref{locrefrac} to evaluate the impact of refraction on the size of spectral features assumes that the atmosphere is well-mixed and that its scale height is constant with altitude. In this Section, we evaluate the impact of a departure from the second assumption by looking in detail at the atmosphere of Jupiter for which we have an in-situ measurement of the temperature-pressure altitude profile from the Galileo probe \citep{Seiff_1998}. We ignore on purpose the impact of clouds as the effects on transmission spectra of an optically thick cloud deck \citep{YB_MS_2017} and their patchy distribution around the planetary limb \citep{L_P_2016} are now well understood, and not the focus of this paper.

\begin{figure*}
\includegraphics[scale=1.0]{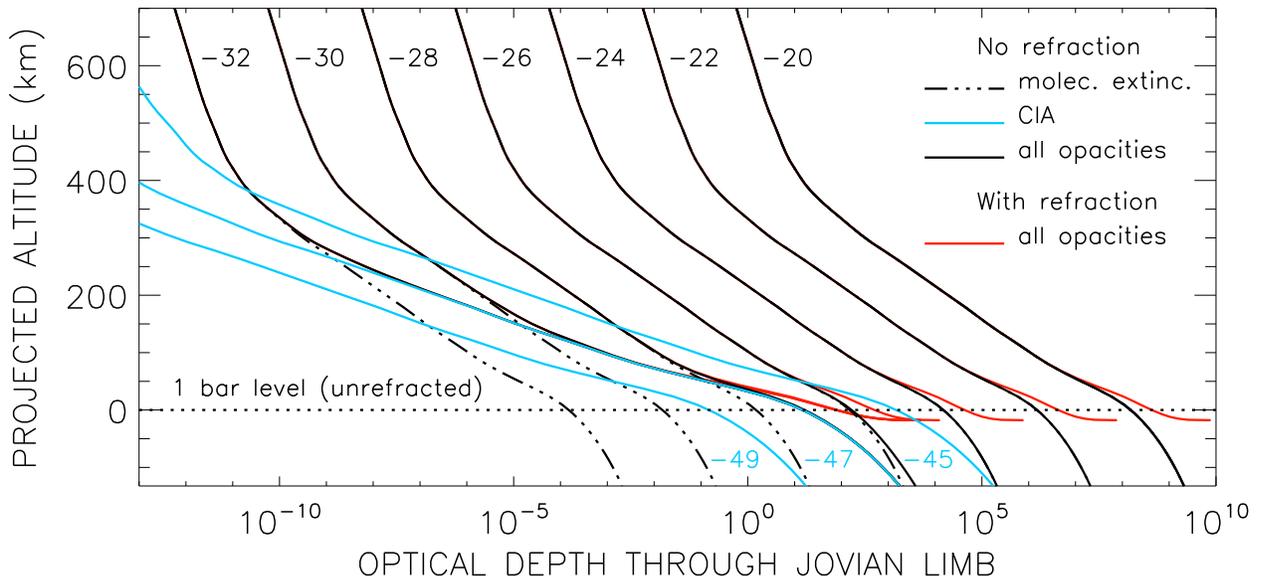}
\caption{Optical depth along a ray traversing the Jovian limb as a function of its projected altitude with respect to the unrefracted 1~bar level for different opacities, assuming a well-mixed atmosphere. Opacities are quantified by the mean atmospheric cross-section, given by $\sigma$ = $10^{X}$~cm$^{2}$~molec$^{-1}$  for atomic and molecular extinction, and k = $3 \times 10^{X}$~cm$^{5}$~molec$^{-2}$ for collision-induced absorption (CIA), where $X$ is the value next to each corresponding curve. The sum of both opacities is only shown for a CIA cross-section value of $3 \times 10^{-47}$~cm$^{5}$~molec$^{-2}$. The figure shows that CIA becomes the dominant opacity only in deeper atmospheric regions, with the transition occurring at lower altitudes as atomic and molecular extinction cross-sections increase. Refraction substantially changes the optical depth altitude profile only in the refractive boundary layer, as described in Sections~\ref{pvsact} through~\ref{probedregion}, as well as \citet{YB_LK_2015}.
\label{fig3}}
\end{figure*}

Since we assume that the atmosphere is well-mixed, the opacity of the atmosphere can be described by a single quantity, the mean extinction cross-section of the atmosphere (see Equ.~\ref{atmcross}), which is constant with altitude. We then obtain the altitude-dependent limb optical depth ($\tau_\sigma$) with
\begin{equation}
\tau_\sigma = N \sigma
\end{equation}
where $N$ is the altitude-dependent column-integrated abundance ($N$) along the limb, computed numerically by MAKEXOSHELL \citep{YB_LK_2015} from the Jovian density altitude profile. We have also upgraded MAKEXOSHELL to compute numerically the column-integrated square of the density ($ \left<N^2 \right>$) along the limb, which we use to compute the altitude-dependent limb CIA optical depth ($\tau_k$) with
\begin{equation}
\tau_k = \left< N^2 \right> k .
\end{equation}
MAKEXOSHELL does these column-integration by first computing the refractivity of the atmosphere from its composition (see Section~\ref{refractivity}) and tracing the ray through the atmosphere outward from a specified grazing altitude \citep{YB_LK_2015}. We treat the non-refractive case by setting the atmospheric refractivity to zero, irrespective of the atmospheric composition. 

Figure~\ref{fig3} shows the limb optical depth altitude profile for atomic and molecular extinctions (triple-dotted dashed line), as well as CIA (light blue line), for a range of mean atmospheric cross sections in the non-refractive case. The sum of both optical depths (black line) is also shown but only for a mean atmospheric CIA cross-section value of $3 \times 10^{-47}$~cm$^{5}$~molec$^{-2}$. Under the well-mixed non-refractive atmosphere assumption, the optical depth traces the density profile of the atmosphere. As expected, we can see from their differing slopes that the relevant scale height for CIA optical depth is about half that of the atomic and molecular extinction optical depth (\citealt{dW_S_2013}; \citealt{YB_MS_2017}). Figure~\ref{fig3} also shows that the atmosphere can be separated in two regions: a lower altitude region where CIA is the dominant source of opacity, and a higher altitude region where atomic and molecular extinctions are. The span of altitudes where the two different opacity sources have about the same optical depth is fairly small. The transition occurs at an altitude where the density $n \approx \sqrt{2} (\sigma/k)$.  As the $(\sigma/k)$ ratio increases, the transition occurs at a higher density, or lower altitude.

The slope of the limb optical depth with altitude is dependent on the scale height, and hence the temperature, of the atmosphere. Shallower slopes occur in the higher temperature regions, or the troposphere (low altitude) and the thermosphere (high altitude). The changing temperature with altitude of Jupiter's atmosphere is responsible for the changing slope with altitude of the optical depth. If Jupiter's temperature was isothermal, the slope would be nearly constant with altitude as gravity only modifies slightly the scale height of the atmosphere with altitude. Refraction modifies the limb optical depth only in the deeper regions of Jupiter's atmosphere (red line), and causes the limb optical depth to increase dramatically with decreasing altitude, essentially mimicking a surface \citep{YB_LK_2015}. Even in the absence of clouds, the effect of Jupiter's refractive boundary layer (also discussed in Sections~\ref{pvsact} through~\ref{probedregion}) hides the temperature signature of its troposphere in an occultation observation geometry, as the slope of the limb optical depth with altitude never becomes shallower at lower altitudes, and instead becomes infinite. 

\begin{figure}
\includegraphics[scale=0.9]{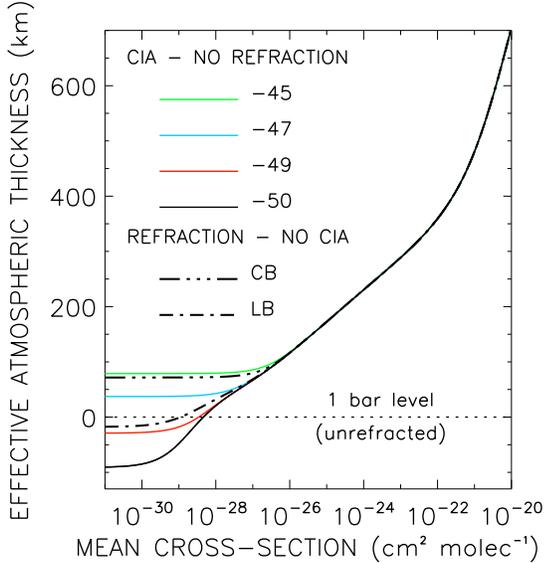}
\caption{Effective atmospheric thickness of a well-mixed Jovian atmosphere with respect to the unrefracted 1~bar pressure level as a function of the mean atomic and molecular extinction cross-section. The solid lines include contribution from CIA, with cross-section values given by k = $3 \times 10^{X}$~cm~$^{5}$ molec$^{-2}$ where $X$ is specified in the legend. The broken lines, which ignore contribution from CIA, show how the refractive critical (CB) and lower (LB) boundaries (see Sections~\ref{lowb} and~\ref{probedregion}) compete with CIA. This figure shows that CIA and refraction only become important in spectral regions where opacity from atomic and molecular extinction is low enough. It also shows the effect of a non-isothermal temperature profile on transmission spectra (see also Fig.~\ref{fig5} and the discussion in Section~\ref{tempeffects}). In an exoplanet transit geometry, Jupiter's critical refractive boundary erases all but the strongest spectral signatures of CIA.
\label{fig4}}
\end{figure}

\begin{figure}
\includegraphics[scale=0.9]{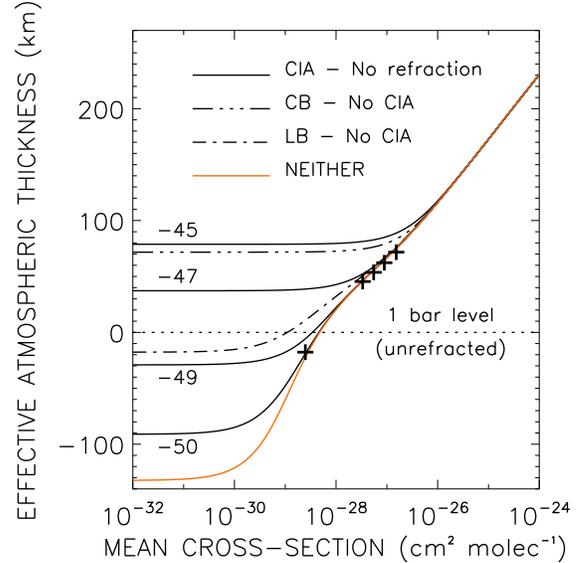}
\caption{Test of the `surface' cross-section concept \citep{YB_MS_2017} for a non-isothermal temperature profile. This figure zooms-in on the lower opacity and lower effective atmospheric thickness corner of Fig.~\ref{fig4} (see its figure caption) with two additional features: the orange line displays the effective atmospheric thickness of the Jovian atmosphere when neither CIA nor refraction are included in the calculations. The near-constant atmospheric thickness at the lowest mean extinction cross-sections is simply due to our computational `surface' at 22~bar, the highest pressure for which the Galileo probe reported temperature measurements \citep{Seiff_1998}. From top to bottom, the top four crosses track the `surface' cross-section associated with Jupiter's critical boundary around a G2, K5, M2 and M5 star, while the bottommost cross tracks the `surface' cross-section associated with the lower refractive boundary. The fact that the crosses follow the orange curve implies that the concept of the `surface' cross-section functions reasonably well even for a non-constant scale height atmosphere.
\label{fig5}}
\end{figure}

From the altitude-dependent limb optical depth, we can compute the effective atmospheric thickness of Jupiter's atmosphere, viewed in an exoplanet transit geometry, using the generalized vertical integration scheme presented by \citet{YB_MS_2017}. Figure~\ref{fig4} shows how the atmospheric thickness of Jupiter's atmosphere, referenced to the unrefracted 1~bar pressure level, changes with the mean atomic and molecular extinction cross-section of the atmosphere. This is shown for several values of the mean CIA cross-section in the non-refractive case (see figure caption), as well as for refractive cases without any CIA contribution to the total optical depth. Figure~\ref{fig5} focusses on the lower regions of the atmosphere with an additional curve (orange line) which has no contribution from either CIA or refraction. The range of values chosen for the mean CIA cross-section spans values of the H$_2$-H$_2$ CIA cross-section measured from 150 to 1000K \citep{Borysow_2002}.

The signature of a non-isothermal temperature structure is manifest by the non-constant slopes of the various lines in Fig.~\ref{fig4} and~\ref{fig5} representing the non-refractive cases (see Section~\ref{tempeffects} for a detailed discussion), except at very low opacities where the effective atmospheric thickness tends to a minimum and the slope tends to zero. The latter is a signature of the underlying constant mean CIA cross-section used in the calculation, which becomes the dominant opacity source at these low mean cross-section values. As the mean CIA cross-section increases, this minimum effective thickness also increases. In the case of the orange curve in Fig.~\ref{fig5} which does not include any CIA opacity, the minimum effective thickness simply coincides with the bottom of the atmosphere defined by the lowest altitude probed by the Galileo probe which we treated as a `surface' below which the atmosphere is completely opaque.

Figures~\ref{fig4} and~\ref{fig5} also show the minimum effective thickness of Jupiter's atmosphere from the refractive continuum for both solar system observations (lower boundary) and exoplanet transits (critical boundary). As with CIA opacity, the effects of refraction are only apparent in spectral regions of low opacity from atomic and molecular sources of extinction. Comparison with published CIA cross-sections (e.g. Borysow, J{\o}rgensen \& Fu 2001; \citealt{Borysow_2002}) reveals that Jupiter's lower boundary (dot-dashed line) hides its deep troposphere and fills-in partly the minima between H$_2$-H$_2$ CIA bands. In contrast, the critical boundary is located almost 90~km above the lower boundary and hides all but the strongest CIA features. 

To test how well the `surface' cross-section concept works for refraction in the case of a non-isothermal atmosphere, we need to see if the effective thickness of a clear non-refractive atmosphere matches that of the refractive continuum when the mean cross-section of the clear atmosphere equals the `surface' cross-section.
We thus compute the critical boundaries, and their corresponding `surface' cross-sections, for a Jupiter-twin orbiting various stars (G2, K5, M2, and M5 spectral class) to see how well these different `surface' cross-sections track the effective thickness of the refractive continuum. To determine the orbital distance of a Jupiter-twin around stars with a spectral class different from our Sun, we used the method described in \citet{YB_LK_2014} with the stellar effective temperatures and radii listed in their Table~1. The method finds the orbital distance which holds the radiation flux at the top of the atmosphere identical to that received by Jupiter in our solar system, assuming that the luminosity of the star is given by that of a perfect blackbody at the specified effective temperature. 

The topmost four crosses in Fig~\ref{fig5} show the surface cross-section associated with the continuum of the critical boundary of a Jupiter-twin orbiting a G2, K5, M2, and M5 star (from top to bottom) while the bottommost cross show the surface cross-section associated with the continuum of the lower boundary. The fact that all the crosses fall very closely onto the orange curve shows that the `surface' cross-section does work to first order, even in a non-isothermal atmosphere, and that it can be used to determine the mean cross-section value below which spectral features may be hidden by the refractive boundary. However, since the scale height is not constant, it is not straightforward to then translate an effective thickness expressed in units of scale height, obtained with Equ.~\ref{spectralmod} and~\ref{spectralmodCIA}, into a transit depth. Instead, this must be computed numerically, as we have done in this Section. 

\subsection{Effects of non-isothermal temperature profiles on transmission spectra}\label{tempeffects}

Figure~\ref{fig4} and~\ref{fig5} are particularly interesting because they illustrate the potential signatures of a non-isothermal atmosphere in an exoplanet's transmission spectrum, as we now demonstrate. Indeed, \citet{Lecavelier_2008} have shown that for a clear well-mixed atmosphere the spectral change in the effective atmospheric thickness ($h$) of the planet is connected to the spectral change of the atmosphere's opacity by
\begin{equation}\label{lecavelier}
\frac{dh}{d\lambda} = H_P \frac{d(\ln\sigma)}{d\lambda} 
\end{equation}
where $\lambda$ is the observed wavelength of the radiation. \citet{YB_MS_2017} have shown analytically that the pressure scale height ($H_P$) should be replaced with the density scale height ($H$) for sources of opacity beside CIA (see also Equ.~\ref{spectralmod}). For CIA, the relevant length scale is $H/2$. 

The density scale height is given by 
\begin{equation}\label{densscale}
\frac{1}{H} = \frac{\overline{m}g}{k_BT} + \frac{\nabla_z T}{T} = \frac{1}{H_P} + \frac{\nabla_z T}{T}
\end{equation}
(Chapter~1 in \citealt{C_H_1987}), where $\overline{m}$ is the mean molecular weight of the atmosphere, $g$ is the gravitational acceleration, $k_B$ is the Boltzmann constant, $\nabla_z T$ is the vertical temperature gradient, and $H_P$ is the pressure scale height. The pressure and density scale height are identical only in isothermal atmospheric regions where $\nabla_z T$ is zero. Atmospheric regions with similar temperatures can have different scale height because the vertical temperature gradient is not necessarily the same. Regions with negative gradients, such as the troposphere, have larger scale height than the pressure scale height, while regions with positive gradients, such as the lower thermosphere, have a smaller scale height than the pressure scale height.

How much larger than the pressure scale height can the density scale height be? The largest negative vertical temperature gradient sustained by an atmosphere, before the atmosphere becomes convectively unstable and restores that gradient, is given by
\begin{equation}\label{maxgrad}
\nabla_z T = -\frac{g}{C_P} = - \left(\frac{\gamma - 1}{\gamma} \right) \frac{\overline{m}g}{k_B} 
\end{equation}
for a dry adiabat. Here, $\gamma = C_P/C_V$ is the ratio of specific heats, where $C_P$ and $C_V$ are the gas specific heat for constant pressure and constant volume, respectively. This means that for a dry adiabat,
\begin{equation}
\frac{\nabla_z T}{T} = - \left(\frac{\gamma - 1}{\gamma}\right) \frac{1}{H_P}
\end{equation}
which yields, after substitution in Equ.~\ref{densscale} and a few algebraic manipulations, the simple result that
\begin{equation}
H = \gamma H_P .
\end{equation}
To first order, $\gamma = 5/3$ for monatomic gases, and $\gamma = 7/5$ for diatomic gases. Hence, the density scale height is significantly larger (about 40\%) than the pressure scale height in precipitation-free convective regions of H$_2$ and N$_2$ atmospheres.

Since the density scale height is independent of wavelength, Equ.~\ref{lecavelier} (with $H_P$ replaced by $H$) can be recast as
\begin{equation}\label{sizefeatures}
\frac{dh}{d(\ln\sigma)} = H 
\end{equation}
which is simply given by the slopes of the various curves in Fig.~\ref{fig4} and~\ref{fig5}. Thus, Fig.~\ref{fig4} and~\ref{fig5} show how the temperature profile in the Jovian atmosphere modulates the size of spectral features in its transmission spectrum. The same change in $\ln\sigma$ maps to a different change in the effective atmospheric thickness depending on the probed atmospheric region. This is readily seen also by inspecting in Fig.~\ref{fig5} how the minimum effective thickness due to CIA changes with CIA opacity. Indeed, a one order of magnitude change from $3 \times 10^{-50}$ to $3 \times 10^{-49}$~cm$^{5}$~molec$^{-2}$ changes the minimum effective thickness of the atmosphere by about 60~km whereas a change by two orders of magnitude from $3 \times 10^{-47}$ to $3 \times 10^{-45}$~cm$^{5}$~molec$^{-2}$ changes it by only 40~km. Except in the troposphere and the lower thermosphere where temperature changes are large, the effects of a changing temperature with altitude are otherwise very subtle, as pointed out by \citet{Barstow_2013} in their study of the effects of the assumption of an isothermal temperature profile on the retrieved abundances of species from exoplanet transmission spectra.

This modulation of the size of spectral features by an atmosphere's temperature profile occurs in all transmission spectra of exoplanets, and the temperature change from the stratosphere to the thermosphere has been detected in a few exoplanets (HD~189733b, HD~209458b) from high spectral resolution transmission spectra centered on the core of atomic lines (\citealt{VidalMadjar_2011}; \citealt{Huitson_2012}; \citealt{VidalMadjar_2013}). The transmission spectrum of a non-isothermal atmosphere is distorted compared to that of an isothermal atmosphere in similar way as our reflections are distorted by fun house mirrors. Regions of larger scale height stretches in the vertical direction spectral features which probe these regions, while regions of smaller scale height compresses them. It is interesting to note that in the case of Jupiter's atmosphere, the lower refractive boundary modifies the slope in the troposphere and masks the signature of its decreasing temperature with altitude in transmission spectra built-up from occultation measurements. Jupiter's critical boundary hides it even further. In hotter atmospheres, refractive boundaries are located deeper (see Section~\ref{jovexo} and \citealt{YB_2016}) and may not be so efficient as Jupiter's atmosphere at hiding the temperature signature of the troposphere, but that could be accomplished by clouds instead.

\section{Exoplanets} \label{exoplanets}

\subsection{Method}\label{exomethod}

Having demonstrated the use of the `surface' cross-section concept as a metric to quantify the effects of refraction on the giant planets in our solar system, we now extend that concept to exoplanets and explore how the `surface' cross-section associated with the critical boundary changes with various parameters. We consider parameters which affect the lensing power of the atmosphere, such as the scale height and refractivity of the planet's atmosphere, as well as the size of the planet, but also the spectral class of the host star which drives the temperature of the atmosphere. We use the same method as discussed in Section~\ref{locrefrac} to determine the location of the critical boundary. The only difference is the input density altitude profile.
The radius of the planet is defined at a reference pressure of 1~bar, while the bottom of the atmosphere is set at a pressure of 500~atm. We spread 80 rays in constant altitude intervals over 30 pressure scale height, of which the latter is evaluated at the bottom of the atmosphere.

As in \citet{YB_2016}, we assume that exoplanets are perfectly spherical and that their atmospheres follow a density altitude profile corresponding to an isothermal temperature-pressure profile at the mean planetary emission temperature ($T_e$), with a gravitational acceleration which varies with altitude. We use a Bond albedo of 0.30, a value similar to Earth and the gas giants in our solar system. The mean planetary emission temperature is identical to the equilibrium temperature ($T_{eq}$), often discussed in the exoplanet literature \citep{Seager_2008}, with an even heat redistribution. This simplifying assumption allows us to relate the planetary temperature to the orbital distance of the planet given the star's spectral type, a necessary step to determine trends in the location of the critical boundary with planetary and stellar temperatures. 

As discussed in Section~\ref{solplanets}, we have upgraded MAKEXOSHELL \citep{YB_LK_2015} with the capability to compute the column-integrated square of the density along the path of a refracted ray, necessary to compute the CIA optical depth with grazing altitude. We have also automated the computation of the location of the various refractive boundaries from the planetary bulk properties (radius, mass, and atmospheric composition), the planetary atmosphere's temperature, and the spectral class of the host star. The location of the refractive boundaries are computed by MAKEXOSHELL in term of their altitude, the density and pressure of the atmosphere, both for the actual and the apparent boundary, which are then also converted into their respective `surface' cross-sections as well. 

\subsection{Jovian exoplanets}\label{jovexo}

\begin{figure}
\includegraphics[scale=0.9]{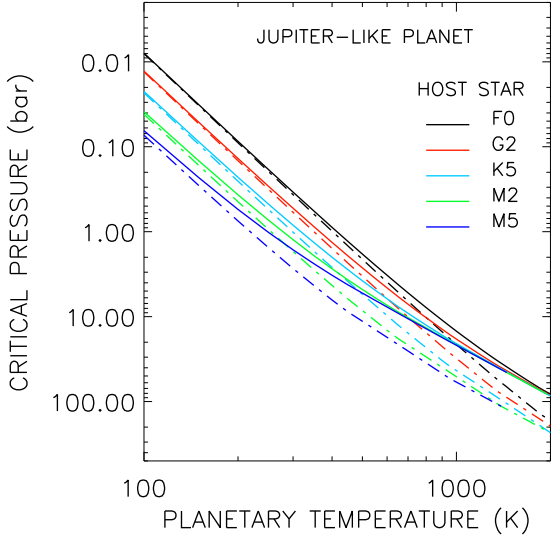}
\caption{Actual (dot-dashed line) and apparent (solid line) critical boundary pressure for a Jupiter-like planet as a function of its isothermal temperature (100 - 2000~K) around different host stars. 
\label{fig6}}
\end{figure}

We first consider an exoplanet with the size, mass, and bulk atmospheric composition of Jupiter as listed in Table~\ref{solsysplanets}. We expand upon the parameter space explored in \citet{YB_2016}. Not only do we add M5 stars to the list of stellar spectral type (F0, G2, K5, M2), but more importantly we consider a much larger number of planetary temperatures than \citet{YB_2016}, which only considered planetary temperatures of 300, 600, and 1200~K. We compute the actual and apparent location of the critical boundary for planetary temperatures ranging from 100 to 2000~K, in increments of 50~K, for a planet orbiting each of the stellar spectral type considered. We show the results in Fig.~\ref{fig6}.

We find that the location of the critical boundary goes to regions of lower pressure -- higher altitudes -- for planets orbiting hotter host stars, as well as lower atmospheric temperatures, in agreement with \citet{YB_2016}. We also find that the differences in the apparent pressure location of the critical boundary, across the range of considered stellar spectral classes, decreases with increasing planetary temperature. This occurs because as the planetary temperature increases, the critical boundary approaches the lower boundary, where not only does a change in the ray deflection requires a smaller altitude change in the grazing radius of the ray, but the effective scale height of the atmosphere also decreases (see Section~\ref{effscl}). This fact had also been demonstrated by \citet{YB_2016}, who showed that for a planetary temperature of 1200~K the location of the refractive continuum was close enough to the lower boundary that its location was almost independent of stellar spectral type.

\begin{figure}
\includegraphics[scale=0.9]{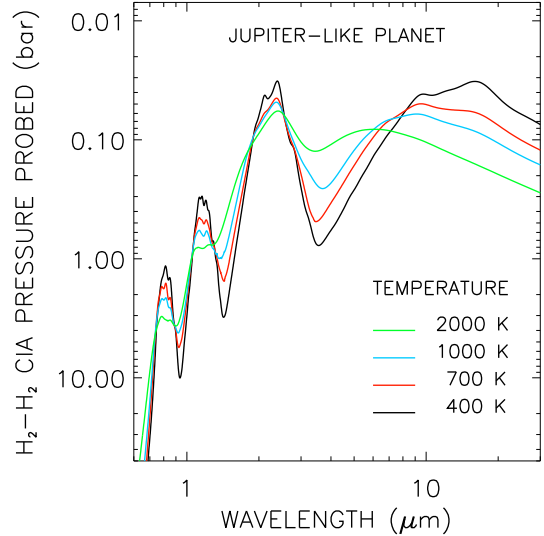}
\caption{Pressure probed by H$_2$ - H$_2$ CIA over the spectral range (0.6 - 30 \micron) covered by all JWST instruments for an isothermal Jovian-like planet at the indicated temperature. We use the CIA cross-sections of \citet{Borysow_2001} and \citet{Borysow_2002} for our calculations.
\label{fig7}}
\end{figure}

What had not been demonstrated by \citet{YB_2016} is how the apparent location of the critical boundary of an isothermal atmosphere with a Jupiter-like composition compares with the pressures probed by CIA, both of which are determined by the bulk composition of the atmosphere. The pressure probed by CIA ($P_{CIA}$) is given by
\begin{equation}\label{ciapress}
P_{CIA} = \left(\frac{G M_P \overline{m}}{\pi}\right)^{1/4} \left(\frac{k_B T}{R_P}\right)^{3/4} \left(\frac{e^{-\gamma_{EM}}}{k} \right)^{1/2}
\end{equation}
for an isothermal atmosphere. Here, the universal constant of gravitation ($G$) and the planetary mass ($M_P$) are the only quantities which we have not yet defined. We derive Equ.~\ref{ciapress} by asking at what pressure does the CIA optical depth along a non-refracted ray equal $e^{-\gamma_{EM}}$ ($\approx 0.56$, \citealt{Lecavelier_2008}; \citealt{dW_S_2013}) given the mean CIA cross-section ($k$, see Equ.~\ref{atmcrossCIA}) of a Jovian atmosphere. The pressure probed by CIA as a function of wavelength is shown in Fig.~\ref{fig7} for the non-refractive case for four different temperatures (400, 700, 1000, and 2000~K). We consider only the most important source of CIA absorption in Jupiter's atmosphere which is from H$_2$-H$_2$, and use the H$_2$-H$_2$ CIA cross-section values published by \citet{Borysow_2001} and \citet{Borysow_2002}. 

A comparison of Fig.~\ref{fig6} and~\ref{fig7} is instructive. The critical boundary of Jovian exoplanets hotter than 1000~K is located at an apparent pressure greater than 10~bar, and CIA absorption features are not impacted by refraction because all CIA features probe lower pressures than this at these temperatures. At 400~K, the critical boundary can mask various parts of the near-infrared CIA features, depending on the spectral class of the host star. Below 200~K, even the size of the strongest CIA features are significantly reduced by refraction. \citet{Robinson_2017} reached similar conclusions from simulations done at three temperatures: 150, 300, and 500~K.

\begin{table*}
 \centering
 \begin{minipage}{175mm}
  \caption{Refractive boundaries of some giant exoplanets assuming a Jovian atmospheric composition}
  \begin{tabular}{@{}ccccccl@{}}
  \hline
  Planet     & T$_e$~\footnote{Assumes a Bond albedo of 0.30} & P$_{c}^*$ (bar) & $\sigma_{c}$ ($10^{-28}$ cm$^2$ molec$^{-1}$)  & k$_{c}$ ($10^{-48}$ cm$^5$ molec$^{-2}$)  &  (r$\nu$/H)$_c$ & Source~\footnote{Reference for planetary and stellar parameters} \\
  \hline
   WASP-17b   &  1615  & 172  & 0.0020  & 0.00037 &  0.45 & \citet{Bento_2014} \\        
   HD 209458b   &  1305 &  62.7 & 0.0098  & 0.0040  &  0.51 & \citet{Knutson_2007}\\   
   WASP-43b   & 1262  & 18.0  & 0.092  & 0.13 &  0.89 & \citet{Hellier_2011} \\   
   HD 189733b  &  1090  & 24.4  &  0.037 & 0.0330  &  0.70 & \citet{Southworth_2010}\\    
   WASP-39b   & 1025  & 69.9 & 0.0053  & 0.0015 &  0.35 & \citet{Maciejewski_2016} \\ 
   HAT-P-26b   & 909   & 80.3  & 0.0075  & 0.0017  & 0.23 & \citet{Hartman_2011} \\        
   Kepler-167e   &  134  & 0.0465  &  9.0  & 500  & 0.051 & \citet{Kipping_2016} \\         
\hline
\end{tabular}\label{exoplanets}
\end{minipage}
\end{table*}

\begin{figure}
\includegraphics[scale=0.9]{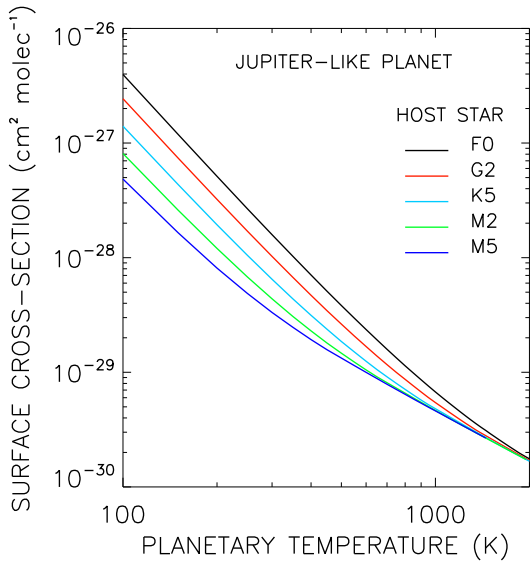}
\caption{`Surface' extinction cross-section ($\sigma_c$) for a Jupiter-like planet as a function of its isothermal temperature (100 - 2000~K) around different host stars. 
\label{fig8}}
\end{figure}

\begin{figure}
\includegraphics[scale=0.9]{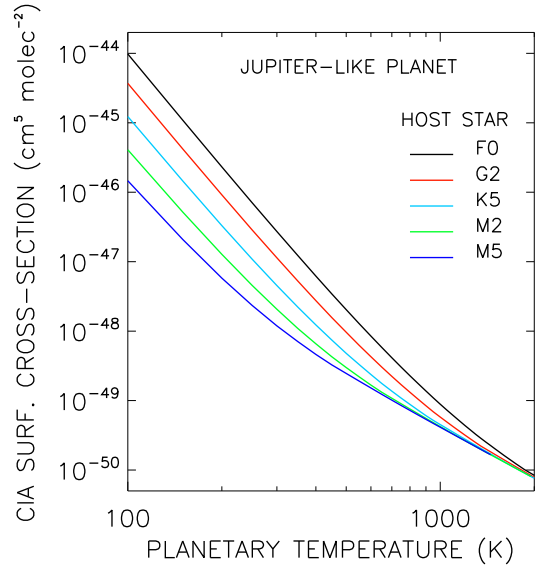}
\caption{`Surface' CIA cross-section ($k_c$) for a Jupiter-like planet as a function of its isothermal temperature (100 - 2000~K) around different host stars. 
\label{fig9}}
\end{figure}

A different but equivalent approach consists in comparing values of the `surface' cross-section to the mean atmospheric 
cross-section (see example in Section~\ref{locrefrac}). Figures~\ref{fig8} and~\ref{fig9} give the corresponding `surface' cross-section to the apparent location of the critical boundary shown in Fig.~\ref{fig6} for atomic and molecular extinction, and CIA, respectively. The advantage of these two figures over Fig.~\ref{fig6}, is that observers can get a quick answer as to whether or not their favorite spectral features are reduced or hidden by refraction by simply looking up extinction cross-sections, factor in the expected abundances of the responsible species, and determine if any part of the computed mean cross-section is lower than the `surface' cross-section. 

We caution the readers that if refractive boundaries are located deep enough that they should lie inside the planet's troposphere where the temperature decreases with altitude, then the refractive boundaries will in fact be located deeper than we predict due to the atmosphere's larger scale height. In such cases, refraction has even less of an effect on transmission spectra than our simple method suggests. However, it is also very likely that refraction did not have much of an impact to begin with, such that our point might be moot. 

Table~\ref{exoplanets} lists the location of the critical boundary for some of the first exoplanets that will be targeted by JWST through the GTO and the ERS (Early Release Science) program, which we rank in order of decreasing temperatures. We chose the properties of the exoplanets by first perusing through the NASA Exoplanet Archive \citep{Akeson_2013}, and selecting sources with a complete coherent set of parameters sufficient to compute the location of the critical boundary. We assume that the bulk composition of their atmosphere is the same as Jupiter's. All of these planets are about the size of Jupiter, except for HAT-P-26b which is Neptune-sized. Although Kepler-167e \citep{Kipping_2016}, a true Jupiter analog, is not on the GTO or ERS target list of JWST, we have included it to emphasize the dramatic effect that the planetary temperature has on the location of the critical boundary. For all the hot exoplanets, the critical boundary is located close to the lower boundary as the factor $(r\nu/H)$ evaluated at the critical boundary is fairly close to one. Indeed, assuming a constant scale height, $(r\nu/H)_c \approx (n_c/n_{lb})$. Among the hot exoplanets, this factor is smallest for HAT-P-26b essentially because of its smaller size. The low values of the CIA `surface' cross-section ($k_c$) implies that spectral features from CIA are not impacted by refraction, and that CIA -- not refraction -- potentially determines the highest pressure that can be probed in the infrared in these hot exoplanets. For the cold Kepler-167e, exactly the opposite occurs.

\begin{figure}
\includegraphics[scale=0.9]{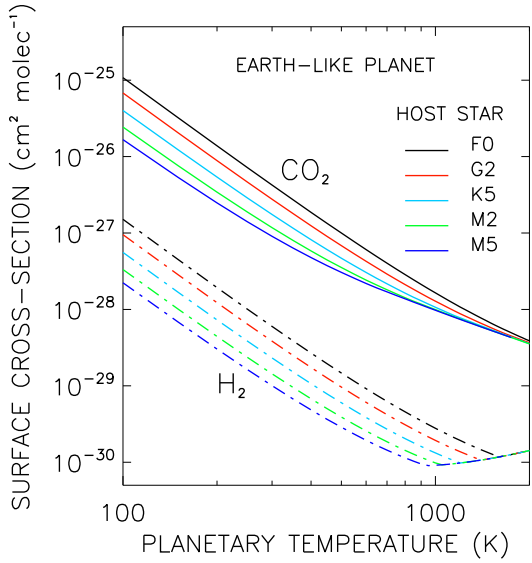}
\caption{`Surface' cross-section ($\sigma_c$) for an Earth-like planet as a function of its isothermal temperature (100 - 2000~K) around different host stars for two different atmospheric composition: H$_2$ (dot-dashed lines), and CO$_2$ (solid lines). The observed deflection of the H$_2$ curves at high temperature simply means that grazing radiation can reach the surface of the planet at a pressure of 500~atm.
\label{fig10}}
\end{figure}

\begin{figure}
\includegraphics[scale=0.9]{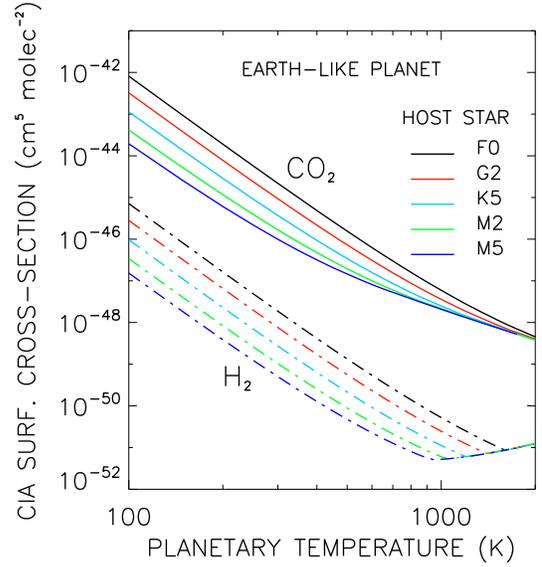}
\caption{`Surface' CIA cross-section ($k_c$) for an Earth-like planet as a function of its isothermal temperature (100 - 2000~K) around different host stars for two different atmospheric composition: H$_2$ (dot-dashed lines), and CO$_2$ (solid lines). The observed deflection of the H$_2$ curves at high temperature simply means that grazing radiation can reach the surface of the planet at a pressure of 500~atm.
\label{fig11}}
\end{figure}
 
\subsection{Terrestrial exoplanets}\label{terrexo}

The importance of refraction in transmission spectra can be a different story for terrestrial exoplanets than for Jovian exoplanets. Not only is the lensing power of an atmosphere lower for smaller planets, but the composition of their atmospheres can be quite different, and the amount of gas making up the atmosphere is finite compared to gas giants. Thus, the actual pressure at the surface may define the largest pressure that can be probed if the atmosphere is thin. However, we focus here on thick atmospheres with a surface pressure of 500~atm and explore how the location of the critical boundary depends on the bulk composition of the atmosphere. We assume that the radius of the planet is located at a reference pressure of 1~bar, as we did for Jovian planets.

\begin{table*}
 \centering
 \begin{minipage}{175mm}
  \caption{Refractive boundaries of some terrestrial exoplanets for different pure atmospheric compositions}
  \begin{tabular}{@{}ccccccccccccccc@{}}
  \hline
   &  &  & \multicolumn{4}{c}{P$_{c}^*$ (bar)} & \multicolumn{4}{c}{$\sigma_{c}$ ($10^{-28}$ cm$^2$ molec$^{-1}$)} & \multicolumn{4}{c}{k$_{c}$ ($10^{-48}$ cm$^5$ molec$^{-2}$)} \\
  Planet     & T$_e$\footnote{Assumes a Bond albedo of 0.30} & (r$\nu$/H)$_c$\footnote{Lowest value is that of a pure H$_2$ atmosphere, while the highest is that of a pure CO$_2$ atmosphere.} & He & H$_2$  & N$_2$  & CO$_2$ & He & H$_2$  & N$_2$  & CO$_2$ & He & H$_2$  & N$_2$  & CO$_2$ \\
  \hline
     GJ1214b        & 550 & 0.22 - 0.70 & 102   & 40.6 & 2.68 & 1.19 & 0.011 & 0.020 & 1.0 & 2.9 & 0.0011 & 0.0053 & 4.2 & 27  \\  
    TRAPPIST-1b  & 366   & 0.12 - 0.45 & 79.2   & 29.7  &  2.63 & 1.24 & 0.018 & 0.036 & 1.3 & 3.5 & 0.0016 & 0.0088 &  3.6 &  20 \\
    TRAPPIST-1c  &  313  & 0.12 - 0.45 & 35.9    & 13.6  &  1.16 & 0.55 & 0.047 & 0.089 & 3.7 & 9.7 & 0.0079 & 0.040 & 19 &  110 \\
    TRAPPIST-1d  &  264  & 0.061 - 0.25 & 33.6    & 12.3  & 1.29  & 0.64  & 0.042 & 0.083 & 2.6 & 6.7 & 0.0064 & 0.035 & 11 & 54 \\
    TRAPPIST-1e  &  230  & 0.055 - 0.23 & 19.1   & 7.0  &  0.73 & 0.36 & 0.063 & 0.12 & 4.1 & 10 & 0.015 & 0.079 & 25 & 130 \\
    TRAPPIST-1f   &  200 & 0.044 - 0.19 & 12.4   & 4.5  & 0.49  & 0.25 & 0.077 & 0.15 & 4.9 & 12 & 0.024 & 0.13 & 40 &190 \\
    TRAPPIST-1g  &  181  & 0.051 - 0.22 & 6.5  & 2.4  & 0.25  & 0.12 & 0.17 & 0.33 & 12 & 29 & 0.092 & 0.48 & 170 & 830 \\       
\hline
\end{tabular}\label{trappist1}
\end{minipage}
\end{table*}

Figures~\ref{fig10} and \ref{fig11} show the `surface' cross-section of the critical boundary for exoplanets with the size and mass as Earth, orbiting M5, M2, K5, G2, and F0 stars, for atomic and molecular extinction, and CIA, respectively. As for Jovian exoplanets, we vary the atmospheric temperature from 100 to 2000~K in 50~K intervals, but we now compute the critical boundary for two different types of atmosphere: a pure H$_2$ and a pure CO$_2$ atmosphere. Both figures show that the `surface' cross-sections of the critical boundary increases by several orders of magnitude from an H$_2$ to a CO$_2$ atmosphere. This occurs because both the refractivity and the mean molecular weight of CO$_2$ are larger than H$_2$. For an H$_2$ atmosphere, the 500~atm surface can be probed if the atmospheric temperature is above 900~K when the planet orbits an M5 star, as indicated by the part of the curve with the positive slope which tracks the apparent location of the surface. The atmospheric temperature at which the critical boundary reaches the surface increases as the temperature of the host star increases. 

Comparison between Earth-sized and Jovian-sized planets of the two types of `surface' cross-section reveal that the `surface' cross-sections of H$_2$ atmospheres are lower for Earth-sized planets. The `surface' cross-sections of a pure CO$_2$ atmosphere of an Earth-sized planet are however higher than that of a Jovian planet because the higher lensing power of a CO$_2$ atmosphere more than compensates for the lower lensing power of a smaller planet. Just as in Jovian planets, the pure CO$_2$ atmosphere of an Earth-sized planet also exhibit a decrease in spread of `surface' cross-sections values between the different host stars with increasing temperatures. This is a telltale sign that the critical boundary approaches the lower boundary well inside the refractive boundary layer at these hotter temperatures.

With the discovery of the TRAPPIST-1 system (\citealt{Gillon_2016}; \citealt{dW_2016}; \citealt{Gillon_2017}) -- a system of at least seven terrestrial planets orbiting an M8 star -- characterization of the composition of atmosphere of terrestrial exoplanets is a reality that is taking shape now. Some of the members of that system will be targeted by JWST during its GTO program. Given the importance of the atmospheric bulk composition on the location of the critical boundary, and the fact that terrestrial planets can accommodate a large diversity of atmospheres, one might wonder how the bulk composition of the atmosphere might affect the integration time required to detect various chemical species. We provide a method to estimate this in the presence of refractive effects for the TRAPPIST-1 planets by exploring how the critical boundary changes with the bulk composition of their atmosphere and expressing it in term of its `surface' cross-section. We use the planetary and stellar parameters which appear in \citet{Gillon_2017}. We do not include TRAPPIST-1h in our calculations because there are no estimate of its mass. We compute the atmospheric temperature in the same manner as the Jovian planets (see Section~\ref{jovexo}), and list them in Table~\ref{trappist1}.

We consider four different gases (H$_2$, He, N$_2$, and CO$_2$), and characterize the different atmospheric compositions in term of the resulting mean molecular weight of the mixture of these gases. We only consider binary mixtures of gases with neighboring mean molecular weight (H$_2$-He, He-N$_2$, and N$_2$-CO$_2$) to avoid possible compositional degeneracies for a given mean molecular weight. Since lighter gases escape planetary atmospheres more readily than heavier gases, this potentially covers different end products of atmosphere stripping scenarios where a terrestrial planet that has undergone little atmospheric loss is composed of primordial H$_2$ and He, whereas it is predominantly composed of He (Hu, Seager, \& Yung 2015), N$_2$ (Earth-like), or CO$_2$ (Venus-like) with increasing loss of the primordial atmosphere. 

\begin{figure}
\includegraphics[scale=0.9]{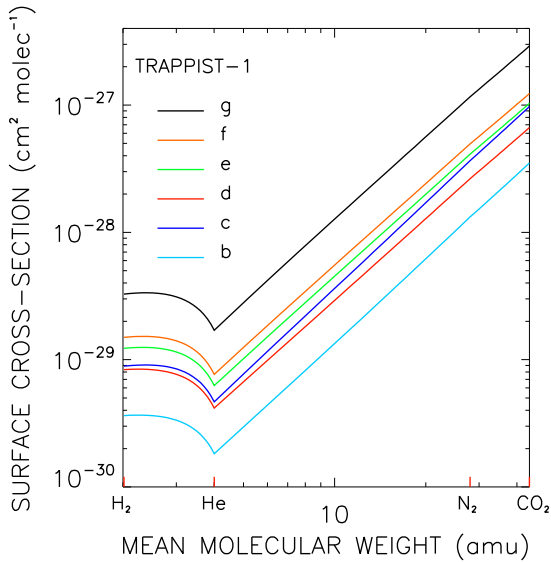}
\caption{`Surface' cross-section ($\sigma_c$) as a function of the mean molecular weight of the atmosphere for the TRAPPIST-1 planets. We assume that the atmosphere is composed of a binary mixture (H$_2$-He, He-N$_2$, and N$_2$-CO$_2$) of only the two gases (identified with red tick marks) bracketing a given mean molecular weight.
\label{fig12}}
\end{figure}

\begin{figure}
\includegraphics[scale=0.9]{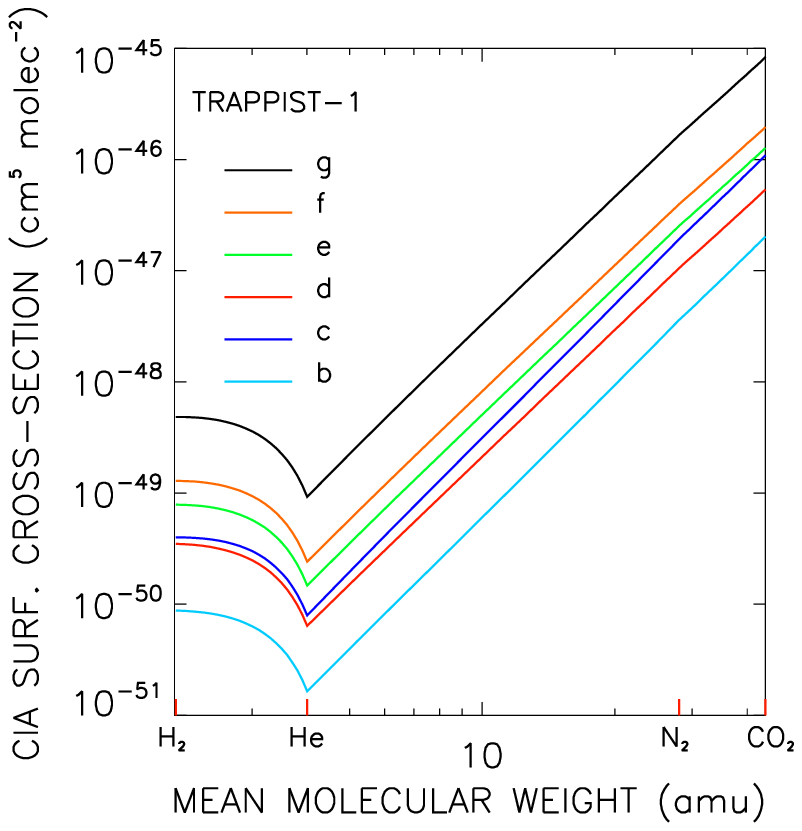}
\caption{`Surface' CIA cross-section ($k_c$) as a function of the mean molecular weight of the atmosphere for the TRAPPIST-1 planets. We assume that the atmosphere is composed of a binary mixture (H$_2$-He, He-N$_2$, and N$_2$-CO$_2$) of only the two gases (identified with red tick marks) bracketing a given mean molecular weight.
\label{fig13}}
\end{figure}

The `surface' cross-sections for atomic and molecular extinction, and for CIA, are shown in Fig.~\ref{fig12} and~\ref{fig13}, respectively, as a function of the mean molecular weight of the atmosphere for the TRAPPIST-1 planets. As expected, the `surface' cross-section generally increases as the orbital distance of the planet increases, because the refractive continuum is located at a higher altitude. The only exception to this trend is between TRAPPIST-1c and -1d, where the much smaller radius of -1d decreases the lensing power of the atmosphere sufficiently that it overcomes the distance dependence. The smallest `surface' cross-section occurs for a pure He atmosphere, because helium's low refractivity decreases its lensing power compared with an H$_2$ atmosphere, in spite of its higher mean molecular weight. A pure CO$_2$ atmosphere has the highest `surface' cross-section. The difference in `surface' cross-sections between the He and CO$_2$ atmosphere amounts to a difference in location of the critical boundary of about $\ln(\sigma_c(CO_2)/\sigma_c(He)) \approx 5$ scale height -- a substantial difference. This implies that as the mean molecular weight of the atmosphere increases, not only do the size of spectral features decrease because of the decreasing scale height, but the refractive continuum shifts to higher altitudes, except in the H$_2$ to He transition, reducing the size of spectral features further. 

The apparent locations of the critical boundary, and their conversion into `surface' cross-sections, for the various pure atmospheres are also summarized in Table~\ref{trappist1}, not only for the TRAPPIST-1 planets, but also for GJ1214b -- an exoplanet which has received a lot of attention -- for which we used the maximum probability value of \citet{AE_2013} for the planetary and stellar parameters. The hotter temperature of GJ1214b generally causes the critical boundary to be located at substantially larger pressures than any of the TRAPPIST-1 planets except for the N$_2$ and CO$_2$ atmosphere. For these atmospheres, the critical boundary is located close to the lower boundary, as shown with a ($r\nu/H$) factor ($\approx n_c/n_{lb}$ for a constant scale height) close to one. The much larger radius of GJ1214b causes that factor to be significantly larger, and the resulting upward shift of the apparent boundary with respect to the actual one causes the apparent pressure of the critical boundary in a CO$_2$ atmosphere to be lower for GJ1214b than for TRAPPIST-1b. Irrespective of the location of the critical boundary, changing the atmosphere of GJ1214 from H$_2$ to CO$_2$ changes the transit depth per scale height, evaluated at the critical boundary, from about 350 to 21~ppm, smaller than the spectral modulation of about 100~ppm observed by \citet{Kreidberg_2014}. However, the lack of correlation between the transmission spectrum and that of a pure CO$_2$ atmosphere favors clouds to explain the fairly flat spectrum.

To illustrate how to use Fig.~\ref{fig12} and~\ref{fig13}, let's consider an example with TRAPPIST-1e (green line) for atomic and molecular extinction (Fig.~\ref{fig12}). Let us suppose that a molecule of interest has a cross-section which ranges from $10^{-25}$ to $10^{-22}$~cm$^{2}$~molec$^{-1}$. If atmospheric models predict an abundance of $10^{-2}$ (mole fraction), then the mean cross-section ranges from $10^{-27}$ to $10^{-24}$~cm$^{2}$~molec$^{-1}$, assuming that the contribution from other sources of opacity is negligible. Since the `surface' cross-section never goes above these values, except barely for a pure CO$_2$ atmosphere, refraction does not reduce the size of that spectral feature, and its spectral modulation is $\ln(10^{-24}/10^{-27}) \approx 7$ scale height. If atmospheric models predict an abundance of $10^{-5}$, then the mean cross-section ranges from $10^{-30}$ to $10^{-27}$~cm$^{2}$~molec$^{-1}$, and the effects of refraction depends on the bulk composition of the atmosphere. For a pure H$_2$ atmosphere, the `surface' cross-section is above $10^{-29}$~cm$^{2}$~molec$^{-1}$, so that its spectral modulation is less than $\ln(10^{-27}/10^{-29}) \approx 4.6$ scale height. In this example, refraction decreases the size of the spectral feature by at least a third from 7 to 4.6 scale heights. For a pure CO$_2$ atmosphere, the `surface' cross-section is larger than the maximum mean cross-section, and the spectral feature is completely hidden by refraction. One can use the same method with CIA on Fig.~\ref{fig13}, except that one must multiply the CIA cross-section by the product of the abundance of the two species responsible for the CIA feature to obtain the mean CIA cross-section. To obtain the spectral modulation expressed in scale height, one must further divide by two the natural logarithm of the ratio of relevant CIA cross-sections (Equ.~\ref{spectralmodCIA}). 

This comparison is potentially complicated by the fact that the extinction cross-sections of gases vary with 
temperature and pressure, which can lead to non-negligible changes in the extrema of opacities, and in the atmospheric pressures probed. Indeed, cross-sections computed at higher pressures and temperatures generally have a lower contrast between the minimum and maximum values of the bands. However, some sources of opacity, such as Rayleigh scattering, hardly change with pressure and temperature, and this comparison can be done easily. One should ideally use cross-sections at the highest possible spectral resolution and at the relevant conditions of temperature and pressure, and determine in the spectral regions of interest how the computed mean cross-section compares to the `surface' cross-section. Since we only consider isothermal atmospheres, it is sufficient for the comparison to use a cross-section close to the equilibrium temperature of the atmosphere for which the `surface' cross-section was computed. We provide these equilibrium temperatures in Tables~\ref{exoplanets} and \ref{trappist1}, and show them in Figures~\ref{fig8} through \ref{fig11}. Since cross-sections have no significant pressure dependence below a transition pressure (see Eq.~\ref{transitp} and the detailed discussion on collisional broadening in the Appendix), cross-sections at low pressures can be approximated by a cross-section computed at zero pressure. Hence, if one wants to know how many scale heights the peak of an absorption features rises above the refractive continuum, one can compare the `surface' cross-section to the mean cross-section computed at zero pressure because most prominent absorption bands probe these low pressures. If instead, one wants to know whether the refractive continuum hides the continuum created by molecular extinction, one should instead use the mean cross-section at the apparent critical pressure. Differences between these two scenarios can yield the spectral modulation present in the transmission spectrum. Indeed, the transmission spectrum of an exoplanet transitions from one computed using zero pressure cross-sections near the pear of absorption bands, to one using cross-sections computed at the apparent critical pressure in the deeper regions.

Choosing a cross-section at the appropriate temperature and pressure is a luxury that is not always available depending on the spectral region or the molecular species of interest. In the ultraviolet, for instance, extinction cross-sections are predominantly measured, rather than computed from first principle, and this only for a limited range of pressures and temperatures. Having at one's disposal a cross-section at a lower spectral resolution, and hence lower contrast, may present some ambiguity in interpretation. Indeed, if the `surface' cross-section is higher than part of said cross-section, it is clear that refraction plays a role in shaping the transmission spectrum. If the `surface' cross-section is always lower, one might wonder if that still would have been the case compared to a higher spectral resolution cross-section which has a higher contrast. In that particular scenario, it is not so clear that refraction has no impact on the spectrum. Even in the infrared where line lists are available, the parameters for collisional broadening are typically known only for self-broadening or air, and only around 300~K. They are poorly known at the high temperatures and for the broadening agents (H$_2$ and He) relevant to hot Jupiters, and HITRAN and EXOMOL -- two line list databases -- have only recently compiled broadening parameters for H$_2$ and He for a few molecules (\citealt{Wilzewski_2016}; \citealt{Barton_2017}). When broadening parameters are not available, one can use our method in an isothermal atmosphere because computed cross-sections cannot change with altitude or pressure.

Whether the atmosphere is probed deep enough that collisional broadening becomes important or not is of secondary concern because our method is not a substitute for a radiative transfer computation, but only meant to help ascertain to first order whether refraction plays a likely role in modifying the transmission spectrum of the exoplanet, given a set of extinction cross-sections and relative abundances of several species of interest. One may simply take a quick look at a published cross-section, and factor in various molecular abundances to get a feel for the range of abundances which can create spectral features above the refractive continuum, or for which spectral features are hidden by refraction.

With this in mind, once the spectral modulation, expressed in scale height ($\Delta h/H$), of a specific spectral feature is determined, one can estimate the corresponding transit depth ($\Delta F/F_{\star}$) associated with this feature with 
\begin{equation}
\left( \frac{\Delta F}{F_\star} \right) = \left( \frac{\Delta h}{H} \right) \left( \frac{2 R_P H}{R_\star^2} \right) . 
\end{equation}
Here, $\Delta F$ is the stellar flux drop caused by the spectral feature, $F_\star$ is the stellar flux, and $R_\star$ is the stellar radius. Fig.~\ref{fig12},~\ref{fig13}, and Table~\ref{trappist1} are thus valuable to get first order effects of refraction for the TRAPPIST-1 planets in order to help determine the exposure time required to detect spectral features 
with various JWST instruments, from an assumed composition of the atmosphere.

\section{Conclusions}\label{conclusion}

We combine refraction theory with the concept of the `surface' cross-section, the result of recent analytical work of the effects of `surfaces' on the transmission spectra of exoplanets \citep{YB_MS_2017} which defines the location of a boundary in opacity space rather than pressure, density, or altitude, to develop a first-order recipe for estimating the effects of the refractive continuum on the size of spectral features. We do this both for atomic and molecular extinction as well as collision-induced absorption. We show, using Jupiter's atmosphere as an example, that the concept of the `surface' cross-section, which defines the value of the mean atmospheric cross-section below which spectral features are severely decreased, works well even for non-isothermal atmospheres. However, unlike for the isothermal case, it is then not straightforward to estimate the resulting flux drop associated with the spectral feature above the refractive continuum because the density scale height changes with altitude. 

Indeed, we demonstrate analytically that the density scale height modulates the change in the effective radius of an exoplanet associated with opacity changes in a well-mixed atmosphere, and that the effects of a non-isothermal atmosphere is to distort the transmission spectrum of an isothermal one: regions of larger scale height stretches in the vertical direction spectral features which probe these regions, while regions of smaller scale height compresses them. In exoplanet atmospheres, these effects are fairly subtle, and only large changes in temperatures in the troposphere and the lower thermosphere could be observed. However, in Jupiter's case, refraction hides the temperature signatures of its troposphere.
 
In comparison to the findings of \citet{Dalba_2015}, we show that differences in the location of the refractive continuum with observational geometry likely lead to almost flat spectra in cold, long-period exoplanet analogs to the Jovian planets in our solar system. Indeed, for both occultations and lunar eclipse observations of solar system planets the lower boundary defines the refractive continuum, while the critical boundary does for exoplanet transmission spectroscopy. 
The difference in location between these two boundaries is at least 4 scale height in altitude for the Jovian planets, which significantly reduces the size of spectral features in exoplanet transmission spectra compared with those reconstructed from solar system observations.

We explore how the location of the critical boundary changes with atmospheric temperature (100 to 2000~K) for both Jupiter-sized and Earth-sized transiting exoplanets orbiting various host stars (M5, M2, K5, G2, F0). An increasing orbital distance of the planet simultaneously decreases the angular size of the host star as seen from the exoplanet, as well as the planetary temperature. Since both these effects increase the altitude of the critical boundary, its location is extremely sensitive to the temperature of the planetary atmosphere. The spread in the location of the critical boundary with the spectral type of the host star decreases with hotter planetary temperature because the critical boundary is located deeper, where the refractive boundary layer of the atmosphere increasingly decreases the effective density scale height of the atmosphere. In Jovian planets with temperatures above 2000~K, the location of the critical boundary is thus essentially independent of stellar spectral type. This is not observed in H$_2$ atmospheres of Earth-sized planets because the lensing power of the atmosphere is smaller for smaller planets. Indeed, radiation can graze a 500~atm surface on an Earth-sized planet when the planetary temperature is above 900~K. If the atmosphere is composed of pure CO$_2$, the resulting increase in the lensing power of the atmosphere more than compensates for the smaller size of the planet, and the same behavior as Jovian planets is observed.

Since the limb optical depth from collision-induced absorption is proportional to the square of the atmospheric density at the grazing radius of the ray, the atmospheric regions where CIA dominates are always deeper than those where atomic and molecular extinction do. Thus, either CIA or refraction defines the deepest regions that can be probed in atmospheric regions of low opacity. In Jovian planets, H$_2$-H$_2$ CIA features are hardly impacted by refraction when the atmosphere is hotter than 1000~K, whereas even the strongest CIA features are significantly reduced by refraction for temperatures colder than 200~K. 

We also explore how the location of the critical boundary changes with the atmospheric composition, quantified in term of the mean molecular weight of a binary mixture (H$_2$-He, He-N$_2$, and N$_2$-CO$_2$), of the planets in the TRAPPIST-1 system. We find that a pure He atmosphere has the lowest critical boundary, a pure CO$_2$ the highest, and that the difference in location amounts to about 5 scale height. As the mean molecular weight of the atmosphere increases above that of helium, not only do absorption features decrease in size due to the decreasing scale height, but the refractive continuum shifts to higher altitude and the spectrum becomes intrinsically flatter.

We provide the location of the refractive boundary for a few exoplanets, including the terrestrial planets of the TRAPPIST-1 system, which will be targeted by JWST in its GTO and ERS program. This helps the scientific community estimate the impact of refraction without the need for complicated radiative transfer calculations, a necessary step to determine the exposure time required to detect spectral features with various JWST instruments. Our results show that the best strategy for JWST observations of the TRAPPIST-1 planets is far from clear. Indeed, the required integration time to detect spectral features of interest varies greatly depending on the composition of the atmosphere. Since we can not know a priori what type of atmosphere to expect on these worlds, do we err on the side of caution by asking for a long integration time possibly sufficient to detect spectral features in a high mean molecular weight atmosphere but also more likely to be denied observation time, or do we err on the side of optimism by planing for an H$_2$ atmosphere and hoping for the best? If we are serious about characterizing the atmospheric composition of another world like our own, we must be ready to pay the observation time toll. Whatever the final adopted strategy is, since the host star of the TRAPPIST-1 system is an extremely cool late-type star (M8), the planetary system is as compact as we can wish for, ideal for minimizing refractive effects, and thus ideal for atmospheric characterization of terrestrial worlds.


\acknowledgments

This research was supported by the appointment of Yan B\'etr\'emieux to the NASA Postdoctoral Program (NPP) at the Jet Propulsion Laboratory, California Institute of Technology. The NPP is administered by Universities Space Research Association (USRA) under contract with NASA. 

This research has made use of the NASA Exoplanet Archive, which is operated by the California Institute of Technology, under contract with the National Aeronautics and Space Administration under the Exoplanet Exploration Program. 

Government sponsorship acknowledged.

\software{MAKEXOSHELL}

\appendix

\section{Impact of collisional broadening on transmission spectra}\label{collbroad}

The simple method discussed in Section~\ref{opacityref}, and applied through a few examples in Sections~\ref{locrefrac} and \ref{terrexo}, to estimate the size of spectral features in the  context of an isothermal atmosphere assumes that absorption cross-sections are independent of altitude. This simple assumption is at the heart of most published discussions on the pressure probed by transmission spectroscopy which draw from the formalism of \citet{Lecavelier_2008}. However, collisional broadening, also known as pressure broadening, can cause absorption cross-sections to vary with pressure or density, as discussed recently at length by \citet{H_M_2016} within the context of exoplanet atmosphere characterization. Indeed, the Voigt profile of a transition line is the convolution of a Gaussian and a Lorentz profile. While the width of the Gaussian profile increases with temperature through Doppler broadening, that of the Lorentz profile not only decreases with temperature but also increases with pressure or density through collisional broadening. The Voigt line profile of a transition thus changes from a Gaussian to a Lorentz profile from low to high densities. As long as the pressure or density is low enough that the Lorentz width is smaller than the Doppler width, the cross-section does not significantly change with altitude in an isothermal atmosphere, and the simple assumption behind our method holds in this Gaussian regime. 

One might wonder at what pressure or density are the two widths equal for molecules for which data is available. This occurs at a Gaussian-to-Lorentz transition pressure ($P_t$) given by
\begin{equation}\label{transitp}
\left( \frac{P_t}{P_{ref}} \right) = \left( \frac{T}{T_{ref}} \right)^\beta \left( \frac{\alpha_D}{\Gamma_{ref}} \right) ,
\end{equation}
which depends on the temperature of the atmosphere ($T$), as well as a collisional broadening temperature scaling exponent ($\beta$) and the Lorentz width ($\Gamma_{ref}$) of the line evaluated at a reference pressure ($P_{ref}$) and temperature ($T_{ref}$) -- typically STP. The last quantity ($\alpha_D$) is the half-width at half-maximum of the Doppler-broadened line, given by 
\begin{equation}
\alpha_D =  \Gamma_D \sqrt{\ln 2}  , 
\end{equation}
where $\Gamma_D$ is its Doppler width. The collisional broadening line parameters $\beta$ and $\Gamma_{ref}$ can be found in line list databases such as HITRAN \citep{Gordon_2017} and EXOMOL \citep{Tennyson_2016}, which have both recently added collisional broadening parameters by H$_2$ and He for a few molecules (\citealt{Wilzewski_2016}; \citealt{Barton_2017}).

\begin{figure}
\includegraphics[scale=0.5]{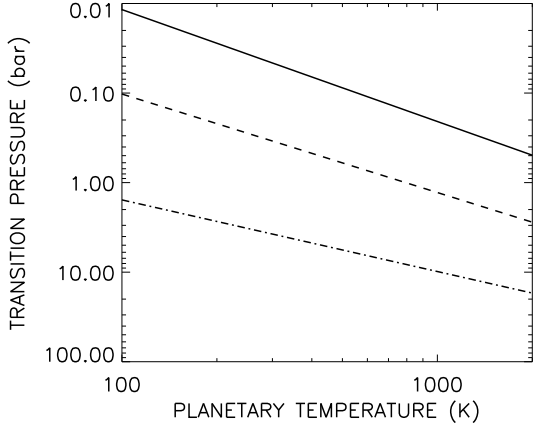}
\caption{Variation of the Gaussian-to-Lorentz transition pressure with atmospheric temperature, computed at a wavelength of 1~$\micron$ for two extreme values of $\Gamma_{ref}m^{1/2}$ from the EXOMOL database, namely from SO$_2$ in H$_2$ (solid) and H$_2$O in He (dot-dashed). The curves of other molecules in that database, which are broadened by H$_2$ and He, lie between them, such as an average curve for H$_2$O in H$_2$ (dashed).
\label{fig14}}
\end{figure}

\begin{figure}
\includegraphics[scale=0.5]{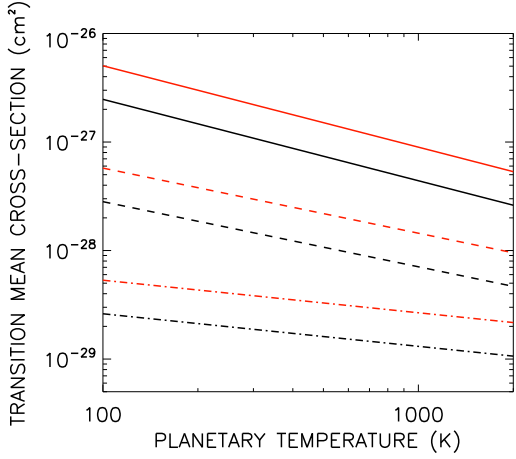}
\caption{Variation of the Gaussian-to-Lorentz transition mean cross-section with atmospheric temperature, computed at a wavelength of 1~$\micron$ mapping to the three curves in Fig.~\ref{fig14}, namely from SO$_2$ in H$_2$ (solid), H$_2$O in H$_2$ (dashed), and H$_2$O in He (dot-dashed). Values are displayed both for a Jupiter-like planet (black), and for an Earth-like planet (red).
\label{fig15}}
\end{figure}

Using the ideal gas law in Equ.~\ref{transitp} to replace pressure with density, and writing out explicitly the expression for the Doppler width, the density ($n_t$) above which the cross-section changes significantly with altitude is then given by
\begin{equation}\label{transitn}
\left( \frac{n_t}{n_{ref}} \right) =  \left( \frac{\sqrt{2 \ln 2}}{\lambda_0 \Gamma_{ref}} \right) \sqrt{\frac{k_B T_{ref}}{mc^2}} \left( \frac{T}{T_{ref}} \right)^{\beta-1/2},
\end{equation}
where $m$ is the mass of the molecule whose line centered on wavelength $\lambda_0$ is broadened. Note that $\Gamma_{ref}$ is typically expressed in cm$^{-1}$ so that the wavelength $\lambda_0$ must be expressed in cm. It is interesting to note that for the few molecules whose collision parameters are tabulated for broadening by H$_2$ and He, $\beta$ typically falls between values of 0.3 and 0.8, which implies that the temperature dependence of the Gaussian-to-Lorentz transition density in Equ.~\ref{transitn} is weak or even non-existent. Hence, the transition density is determined predominantly by the product of the wavelength, the reference Lorentz line width, and the square root of the mass of the broadened species ($ \lambda_0 \Gamma_{ref}m^{1/2}$). As wavelength increases, the transition density decreases and collisional broadening is important over a larger fraction of the atmosphere. The other two quantities depend on the chemical species whose lines are broadened. The expression for $(P_t/P_{ref})$ is identical to that of $(n_t/n_{ref})$ in Eq.~\ref{transitn} except for the exponent above the temperature ratio which is $\beta+1/2$.

Fig.~\ref{fig14} shows the upper (H$_2$O in He) and lower (SO$_2$ in H$_2$) boundaries of the range of transition pressures at 1~$\micron$ that molecules in the EXOMOL database, which are broadened by H$_2$ and He, can have. These boundaries are set by the combination of molecule and broadener which have the lowest and highest value of the $\Gamma_{ref}m^{1/2}$ product, i.e. 0.017~amu$^{1/2}$~cm$^{-1}$ and 1.44~amu$^{1/2}$~cm$^{-1}$ respectively. The intermediate curve shows a typical result for H$_2$O in H$_2$ with a $\Gamma_{ref}m^{1/2}$ product of 0.19~amu$^{1/2}$~cm$^{-1}$.  As $\Gamma_{ref}$ and $\beta$ varies from line to line, it is understood that cross-sections can have different transition pressure in different spectral regions. Nevertheless, it is instructive to see the envelope of pressures which describe the possible locations of this transition region. Furthermore, since the transition pressure is inversely proportional to wavelength, the corresponding transition pressures at 10~$\micron$ are 10 times smaller than those at 1~$\micron$. Comparing our results to those of \citet{H_M_2016}, it is puzzling that the minimum transition pressure which we compute is about 10~mbar at 100~K, which is several orders of magnitude greater than displayed in their Fig.~2. However, \citet{H_M_2016} do not specify either the wavelength, the molecule, or the linewidth for which their transition pressure was computed. Furthermore, the bottom panel of their Fig.~9 shows that the cross-section at 5~$\micron$ of H$_2$O in H$_2$ at a temperature of 1000~K starts changing significantly with pressures somewhere between 0.01 and 0.1~atm. Although this is two orders of magnitude higher than the value of about 0.0001~atm shown in their Fig.~2, it is however consistent with the range of values displayed in our Fig.~\ref{fig14}, factoring the factor of five from the wavelength dependence. It would appear that the transition pressures in Fig.~2 in \citet{H_M_2016} are orders of magnitude too small, and inflate the importance of collisional broadening in exoplanet transmission spectra.

Our calculations imply that collisional broadening in primordial atmospheres of hot exoplanets may only be important in transmission spectroscopy in the wings of prominent absorption bands because the Gaussian-to-Lorentz transition occurs at higher pressures than are typically probed by the peak of these bands. Indeed, one can compute the mean cross-section of a homogenous isothermal atmosphere which probes this Gaussian-to-Lorentz transition region by determining where the optical depth is 0.561 along the limb. This transition mean cross-section ($\sigma_t$) is given by
\begin{equation}\label{transitsigma}
\sigma_t  = \frac{e^{-\gamma_{EM}}}{2 \sqrt{\pi \ln 2}} \sqrt{\frac{GM_P}{R_P^3}} \left( \frac{m^*c}{k_B T_{ref}} \right)\left( \frac{\lambda_0 \Gamma_{ref}}{L_0} \right) \left( \frac{T}{T_{ref}} \right)^{-\beta},
\end{equation}
where 
\begin{equation}
m^* = \sqrt{m\overline{m}} .
\end{equation}
Here, $\overline{m}$ is the mean molecular weight of the atmosphere, $G$ is the gravitational constant, and $R_P$ and $M_P$ are the radius and the mass of the planet, respectively. Unlike the transition density or pressure, the transition mean cross-section depends on the bulk atmospheric properties, as well as the radius and mass of the exoplanet. Our simple approximation of a constant mean cross-section with altitude across the probed atmospheric regions holds in spectral regions where $\sigma > \sigma_t$. Fig.~\ref{fig15} shows this transition mean cross-section corresponding to the curves in Fig~\ref{fig14} for exoplanets with the same radius and mass as Jupiter and Earth. 

A quick comparison between our calculation for H$_2$O in H$_2$ with the cross-section shown in Fig.~9 of \citet{H_M_2016} is quite instructive. For a 1000~K atmosphere of a Jupiter-sized planet, the prominent bands of the H$_2$O cross-section always probe lower pressures than the transition pressure if the water relative abundance is greater than 10$^{-5}$. Indeed, the low-pressure peak of those bands have a cross-section value of at least 10$^{-21}$~cm$^2$, which combined with a relative abundance of 10$^{-5}$ results in a minimum mean cross-section of 10$^{-26}$~cm$^2$. This value is greater than the upper envelope of possible transition values from 4$\times$10$^{-28}$~cm$^2$ at 1~$\micron$ shown in Fig.~\ref{fig15} (solid black line) to 8$\times$10$^{-27}$ at 20~$\micron$. It is even greater than the typical transition mean cross-section of water in H$_2$ (dashed black line) of about 7$\times$10$^{-29} - $1.4$\times$10$^{-27}$~cm$^2$ in that same spectral range. 

One can also consider the inverse problem and ask what part of the low-pressure cross-section contributing to the exoplanet transmission spectrum is impacted by collisional broadening given the relative abundance of the molecule considered. Considering the above typical transition mean cross-section of water in H$_2$ with this same relative abundance of 10$^{-5}$ implies that cross-section values below the line connecting 7$\times$10$^{-24}$ at 1~$\micron$ to 1.4$\times$10$^{-22}$~cm$^2$ at 20~$\micron$, on a $\log(\sigma)$ versus $\log(\lambda)$ plot, are impacted by collisional broadening. The entire line moves downward with increasing relative abundance, and collisional broadening becomes less important. The resulting transmission spectrum resemble the logarithm of the low-pressure cross-sections above this line, while it progressively resembles the logarithm of the cross-sections at higher pressures at lower altitudes. 
Probing these deeper regions is however possible only if they are not hidden by the absorption band of other molecules. When faced with complex atmospheres composed of many molecular species, this is unlikely to happen over a significant fraction of the observed spectrum.


\begin{thebibliography}{}

\bibitem[\protect\citeauthoryear{Akeson et al.}{2013}]{Akeson_2013} Akeson R. L., Chen X., Ciardi D.,
Crane M., Good J., et al., 2013, PASP, 125, 989A
\bibitem[\protect\citeauthoryear{Anglada-Escud\'e et al.}{2013}]{AE_2013} Anglada-Escud\'e G., Rojas-Ayala B., Boss A. P., Weinberger A. J., Lloyd J. P., 2013, A\&A, 551, A48
\bibitem[\protect\citeauthoryear{Arney et al.}{2017}]{Arney_2017} Arney G. N., et al., 2017, ApJ, 836, 49
\bibitem[\protect\citeauthoryear{Arnold et al.}{2014}]{Arnold_2014} Arnold L., Ehrenreich D., Vidal-Madjar A.,
Dumusque X., Nitschelm C., Querel R. R., et al., 2014, A\&A, 564, A58
\bibitem[\protect\citeauthoryear{Barstow \& Irwin}{2016}]{B_I_2016} Barstow J. K., Irwin P. G. J., 2016, MNRAS, 461, L92
\bibitem[\protect\citeauthoryear{Barstow et al.}{2013}]{Barstow_2013} Barstow J. K., Aigrain S.,  Irwin P. G. J., Bowles N., Fletcher L. N., Lee J.-M., 2013, MNRAS, 430, 1188
\bibitem[\protect\citeauthoryear{Barstow et al.}{2015}]{Barstow_2015} Barstow J. K., Aigrain S.,  Irwin P. G. J., Kendrew S., Fletcher L. N., 2015, MNRAS, 448, 2546
\bibitem[\protect\citeauthoryear{Barstow et al.}{2016}]{Barstow_2016} Barstow J. K., Aigrain S.,  Irwin P. G. J., Kendrew S., Fletcher L. N., 2016, MNRAS, 458, 2657
\bibitem[\protect\citeauthoryear{Barstow et al.}{2017}]{Barstow_2017} Barstow J. K., Aigrain S.,  Irwin P. G. J., Sing D. K., 2017, ApJ, 834, 50
\bibitem[\protect\citeauthoryear{Barton et al.}{2017}]{Barton_2017} Barton E. J., Hill C., Czurylo M., Li H. Y., Hyslop A., Yurchenko S. N., Tennyson J., 2017, JQSRT, 203, 490
\bibitem[\protect\citeauthoryear{Baum \& Code}{1953}]{B_C_1953} Baum W. A., Code A. D., 1953, AJ, 
58, 108
\bibitem[\protect\citeauthoryear{Beichman et al.}{2014}]{Beichman_2014} Beichman C., et al., 2014, PASP, 126, 1134
\bibitem[\protect\citeauthoryear{Benneke \& Seager}{2012}]{B_S_2012} Benneke B., Seager S., 2012, ApJ, 753, 100
\bibitem[\protect\citeauthoryear{Bento et al.}{2014}]{Bento_2014} Bento J., et al., 2014, MNRAS, 437, 1511
\bibitem[\protect\citeauthoryear{Berta et al.}{2012}]{Berta_2012} Berta Z. K., et al., 2012, ApJ, 747, 35
\bibitem[\protect\citeauthoryear{B\'etr\'emieux}{2016}]{YB_2016} B\'etr\'emieux Y., 2016, MNRAS, 456, 4051
\bibitem[\protect\citeauthoryear{B\'etr\'emieux \& Kaltenegger}{2013}]{YB_LK_2013} B\'etr\'emieux Y., Kaltenegger L., 2013, ApJ, 772, L31
\bibitem[\protect\citeauthoryear{B\'etr\'emieux \& Kaltenegger}{2014}]{YB_LK_2014} B\'etr\'emieux Y., Kaltenegger L., 2014, ApJ, 791, 7
\bibitem[\protect\citeauthoryear{B\'etr\'emieux \& Kaltenegger}{2015}]{YB_LK_2015} B\'etr\'emieux Y., Kaltenegger L., 2015, MNRAS, 451, 1268
\bibitem[\protect\citeauthoryear{B\'etr\'emieux \& Swain}{2017}]{YB_MS_2017} B\'etr\'emieux Y., Swain M. R., 2017, MNRAS, 467, 2834
\bibitem[\protect\citeauthoryear{Borysow}{2002}]{Borysow_2002} Borysow A., 2002, A\&A 390, 779
\bibitem[\protect\citeauthoryear{Borysow et al.}{2001}]{Borysow_2001} Borysow A., J{\o}rgensen U. G., Fu Y., 2001, JQSRT, 68, 235 
\bibitem[\protect\citeauthoryear{Brown}{2001}]{Brown_2001} Brown T. M., 2001, ApJ, 553, 1006
\bibitem[\protect\citeauthoryear{Burrows}{2014}]{Burrows_2014} Burrows A., 2014, PNAS 111, 12603
\bibitem[\protect\citeauthoryear{Chamberlain \& Hunten}{1987}]{C_H_1987} Chamberlain J. W., Hunten D. M., 1987, Theory of planetary atmospheres, 2nd Edition, Academic Press, Inc., San Diego, CA
\bibitem[\protect\citeauthoryear{Conrath \& Gautier}{2000}]{C_G_2000} Conrath B. J., Gautier D., 2000, Icarus, 144, 124
\bibitem[\protect\citeauthoryear{Dalba et al.}{2015}]{Dalba_2015} Dalba P. A., Muirhead P. S., Fortney J. J., Hedman M. M., Nicholson P. D., Veyette M. J., 2015, ApJ, 814, 154
\bibitem[\protect\citeauthoryear{Deming et al.}{2009}]{Deming_2009} Deming D., et al., 2009, PASP, 121, 952
\bibitem[\protect\citeauthoryear{de Wit \& Seager}{2013}]{dW_S_2013} de Wit J., Seager S., 2013, Science, 342, 1473
\bibitem[\protect\citeauthoryear{de Wit et al.}{2016}]{dW_2016} de Wit J., Wakeford H. R., Gillon M., et al., 2016, Nature, 537, 69
\bibitem[\protect\citeauthoryear{Evans et al.}{2016}]{Evans_2016} Evans T. M., et al., 2016, ApJ, 822, L4
\bibitem[\protect\citeauthoryear{Fortney}{2005}]{Fortney_2005} Fortney J.J., 2005, MNRAS, 364, 649
\bibitem[\protect\citeauthoryear{Garc\'ia Mu\~{n}oz \& Mills}{2012}]{Munoz_Mills_2012} Garc\'ia Mu\~{n}oz A., Mills F. P., 2012, A\&A, 547, A22
\bibitem[\protect\citeauthoryear{Garc\'ia Mu\~{n}oz et al.}{2012}]{Munoz_2012} Garc\'ia Mu\~{n}oz A., 
Zapatero Osorio M. R., Barrena R., Monta\~{n}\'{e}s-Rodr\'{i}guez P., Mart\'{i}n E. L., Pall\'{e} E., 2012, 
ApJ, 755, 103
\bibitem[\protect\citeauthoryear{Gibson}{2014}]{Gibson_2014} Gibson N. P., 2014, MNRAS, 445, 340
\bibitem[\protect\citeauthoryear{Gillon et al.}{2016}]{Gillon_2016} Gillon M., et al., 2016, Nature, 533, 221
\bibitem[\protect\citeauthoryear{Gillon et al.}{2017}]{Gillon_2017} Gillon M., et al., 2017, Nature, 542, 456
\bibitem[\protect\citeauthoryear{Gordon et al.}{2017}]{Gordon_2017} Gordon I. E., et al., 2017, JQSRT, 203, 3
\bibitem[\protect\citeauthoryear{Greene et al.}{2016}]{Greene_2016} Greene T. P., Line M. R., Montero C., Fortney J. J., Lustig-Yaeger J., Luther K., 2016, ApJ, 817, 17
\bibitem[\protect\citeauthoryear{Hartman et al.}{2011}]{Hartman_2011} Hartman J. D., et al., 2011, ApJ, 728, 138
\bibitem[\protect\citeauthoryear{Hedges \& Madhusudhan}{2016}]{H_M_2016} Hedges C., Madhusudhan N., 2016, MNRAS, 458, 1427
\bibitem[\protect\citeauthoryear{Hellier et al.}{2011}]{Hellier_2011} Hellier C., et al., 2011, A\&A, 535, L7
\bibitem[\protect\citeauthoryear{Hu et al.}{2015}]{Hu_2015} Hu R., Seager S., Yung Y. L., 2015, ApJ, 807, 8
\bibitem[\protect\citeauthoryear{Hubbard et al.}{2001}]{Hubbard_2001} Hubbard W. B., Fortney J. J., Lunine J. I., Burrows A., Sudarsky D., Pinto P., 2001, ApJ, 560, 413
\bibitem[\protect\citeauthoryear{Huitson et al.}{2012}]{Huitson_2012} Huitson C. M., et al., 2012, MNRAS, 422, 2477
\bibitem[\protect\citeauthoryear{Irwin et al.}{2008}]{Irwin_2008} Irwin P. G. J. et al., 2008, JQSRT, 109, 1136
\bibitem[\protect\citeauthoryear{Iyer et al.}{2016}]{Iyer_2016} Iyer A. R., Swain M. R., Zellem R. T., Line M. R., Roudier G., Rocha G., Livingston J. H., 2016, ApJ, 823, 109
\bibitem[\protect\citeauthoryear{Kipping et al.}{2016}]{Kipping_2016} Kipping D. M., et al., 2016, ApJ, 820, 112
\bibitem[\protect\citeauthoryear{Knutson et al.}{2007}]{Knutson_2007} Knutson H. A., Charbonneau D., Noyes R. W., Brown T. M., Gilliland R. L., 2007, ApJ, 655, 564
\bibitem[\protect\citeauthoryear{Kreidberg et al.}{2014}]{Kreidberg_2014} Kreidberg L., et al., 2014, Nature, 69, 505
\bibitem[\protect\citeauthoryear{Lecavelier des Etangs et al.}{2008}]{Lecavelier_2008} Lecavelier des Etangs A., Pont F., Vidal-Madjar A., Sing D., 2008, A\&A, 481, L83
\bibitem[\protect\citeauthoryear{L\'eger et al.}{2004}]{Leger_2004} L\'eger A., Selsis F., Sotin C., et al. 2004, Icarus, 169, 499
\bibitem[\protect\citeauthoryear{Lindal}{1992}]{Lindal_1992} Lindal G. F., 1992, Astronomical Journal, 103, 967
\bibitem[\protect\citeauthoryear{Lindal et al.}{1985}]{Lindal_1985} Lindal G. F., Sweetnam D. N., Eshleman V. R., 1985, 
Astronomical Journal, 90, 1136
\bibitem[\protect\citeauthoryear{Lindal et al.}{1987}]{Lindal_1987} Lindal G. F., Lyons J. R., Sweetnam D. N., Eshleman V. R., Hinson D. P., Tyler G. L., 1987, JGR, 92, 14987 
\bibitem[\protect\citeauthoryear{Line \& Parmentier}{2016}]{L_P_2016} Line M. R., Parmentier V., 2016, ApJ, 820, 78
\bibitem[\protect\citeauthoryear{Line et al.}{2013}]{Line_2013} Line M. R., et al., 2013, ApJ, 775, 137
\bibitem[\protect\citeauthoryear{Lodders \& Fegley, Jr.}{1998}]{L_F_1998} Lodders K., Fegley Jr. B., 1998, 
The Planetary Scientist's Companion, Oxford Univ. Press, New York
\bibitem[\protect\citeauthoryear{Maciejewski et al.}{2016}]{Maciejewski_2016} Maciejewski G., et al., 2016, Acta Astronomica, 66, 55
\bibitem[\protect\citeauthoryear{Madhusudhan \& Seager}{2009}]{M_S_2009} Madhusudhan N., Seager S., 2009, ApJ, 707, 24
\bibitem[\protect\citeauthoryear{McCullough et al.}{2014}]{McCullough_2014} McCullough P. R., Crouzet N., 
Demming D., Madhusudhan N., 2014, ApJ, 791, 55
\bibitem[\protect\citeauthoryear{MacDonald \& Madhusudhan}{2017}]{M_M_2017} MacDonald R. J., Madhusudhan N., 2017, MNRAS, 469, 1979
\bibitem[\protect\citeauthoryear{Misra, Meadows \& Crisp}{2014}]{MMC_2014} Misra A., Meadows V.,
Crisp D., 2014, ApJ, 792, 61
\bibitem[\protect\citeauthoryear{Molli\`ere et al.}{2017}]{Molliere_2017} Molli\`ere P., van Boekel R.,
Bouwman J., Henning T., Lagage P.-O., Min M., 2017, A\&A, 600, A10
\bibitem[\protect\citeauthoryear{Pall\'{e} et al.}{2009}]{Palle_2009} Pall\'{e} E., Zapatero Osorio M. R.,
Barrena R., Monta\~{n}\'{e}s-Rodr\'{i}guez P., Mart\'{i}n E. L., 2009, Nature, 459, 814
\bibitem[\protect\citeauthoryear{Phinney \& Anderson}{1968}]{P_A_1968} Phinney R. A., Anderson D. L.,
1968, JGR, 73, 1819
\bibitem[\protect\citeauthoryear{Pont et al.}{2008}]{Pont_2008} Pont F., Knutson H., Gilliland R. L., Moutou C., Charbonneau D., 2008, MNRAS, 385, 109
\bibitem[\protect\citeauthoryear{Pont et al.}{2013}]{Pont_2013} Pont F., Sing D. K., Gibson N. P., Aigrain S., Henry G., Husnoo N., 2013, MNRAS, 423, 2917
\bibitem[\protect\citeauthoryear{Robinson et al.}{2014}]{Robinson_2014} Robinson T. D., Maltagliati L., Marley M. S., Fortney J. J., 2014, PNAS, 111, 9042
\bibitem[\protect\citeauthoryear{Robinson et al.}{2017}]{Robinson_2017} Robinson T. D., Fortney J. J., Hubbard W. B., 2017, ApJ, 850, 128
\bibitem[\protect\citeauthoryear{Seager}{2008}]{Seager_2008} Seager S., 2008, Space Sci. Rev., 135, 345
\bibitem[\protect\citeauthoryear{Seiff et al.}{1985}]{Seiff_1985} Seiff A., Schofield J. T., Kliore A. J.,
Taylor F. W., Limaye  S. S., et al., 1985, Adv. Space Res., 5, 3
\bibitem[\protect\citeauthoryear{Seiff et al.}{1998}]{Seiff_1998} Seiff A., et al., 1998, JGR, 103, 22857
\bibitem[\protect\citeauthoryear{Sidis \& Sari}{2010}]{Sidis_Sari_2010} Sidis O., Sari, R., 2010, ApJ, 720, 904
\bibitem[\protect\citeauthoryear{Sing et al.}{2011}]{Sing_2011} Sing D. K., et al., 2011, MNRAS, 416, 1443
\bibitem[\protect\citeauthoryear{Sing et al.}{2015}]{Sing_2015} Sing D. K., et al., 2015, MNRAS, 446, 2428
\bibitem[\protect\citeauthoryear{Sing et al.}{2016}]{Sing_2016} Sing D. K., et al., 2016, Nature, 529, 59
\bibitem[\protect\citeauthoryear{Smith \& Hunten}{1990}]{S_H_1990} Smith G. R., Hunten D. M., 1990, Rev. Geophys., 28, 117
\bibitem[\protect\citeauthoryear{Southworth}{2010}]{Southworth_2010} Southworth J., 2010, MNRAS, 408, 1689
\bibitem[\protect\citeauthoryear{Stevenson}{2016}]{Stevenson_2016} Stevenson K. B., 2016, ApJ, 817, L16
\bibitem[\protect\citeauthoryear{Stevenson et al.}{2016}]{S_JWST_2016} Stevenson K. B., et al., 2016, PASP, 128, 1
\bibitem[\protect\citeauthoryear{Swain et al.}{2008}]{Swain_2008} Swain M. R., Vasisht G., Tinetti G., 2008, Nature, 452, 329
\bibitem[\protect\citeauthoryear{Tennyson et al.}{2016}]{Tennyson_2016} Tennyson J., et al., 2016, Journal of Molecular Spectroscopy, 327, 73
\bibitem[\protect\citeauthoryear{Tinetti et al.}{2007}]{Tinetti_2007} Tinetti G., et al., 2007, Nature, 448, 169
\bibitem[\protect\citeauthoryear{Vidal-Madjar et al.}{2010}]{VidalMadjar_2010} Vidal-Madjar A., et al., 2010, A\&A, 523, A57
\bibitem[\protect\citeauthoryear{Vidal-Madjar et al.}{2011}]{VidalMadjar_2011} Vidal-Madjar A., et al., 2011, A\&A, 527, A110
\bibitem[\protect\citeauthoryear{Vidal-Madjar et al.}{2013}]{VidalMadjar_2013} Vidal-Madjar A., et al., 2013, A\&A, 560, A64
\bibitem[\protect\citeauthoryear{Waldmann et al.}{2015}]{Waldmann_2015} Waldmann I. P., Tinetti G., Rocchetto M., Barton E. J., Yurchenko S. N., Tennyson J., 2015, ApJ, 802, 107
\bibitem[\protect\citeauthoryear{Wilzewski et al.}{2016}]{Wilzewski_2016} Wilzewski J. S., Gordon I. E., Kochanov R. V., Hill C., Rothman L. S., 2016, JQSRT, 168, 193
\bibitem[\protect\citeauthoryear{Yan et al.}{2015}]{Yan_2015} Yan F., Fosbury R. A. E., Petr-Gotzens M. G., Zhao G., Wang W., Wang L., Liu Y., Pall\'e E., 2015, IJA, 14, 255
\\

\end{thebibliography}
\end{document}